\documentclass[prd,aps,twocolumn,superscriptaddress,amsmath,amssymb,floatfix]{revtex4-2}

\usepackage{lineno}
\usepackage{soul}
\usepackage{subfigure}
\usepackage[export]{adjustbox}
\RequirePackage{snapshot}
\usepackage[colorlinks=true,citecolor=blue,filecolor=blue,linkcolor=blue,urlcolor=blue]{hyperref}

\usepackage{graphicx}           
\usepackage{array}              
\usepackage{placeins}           
\usepackage[dvipsnames]{xcolor}  
\usepackage{tikz}               
\usetikzlibrary{shapes, decorations.pathreplacing}  

\usepackage{amsmath}            
\usepackage{amssymb}            
\usepackage{siunitx}            
\usepackage{xspace}             
\usepackage{bm}                 
\usepackage{ifthen}

\usepackage[T1]{fontenc}
\usepackage[utf8]{inputenc}
\usepackage[english]{babel}
\usepackage{csquotes}           

\usepackage{lipsum}             
\usepackage{verbatim}           
\usepackage{pgf}                
\usepackage{xcolor}             
\usepackage[utf8]{inputenc}     
\usepackage{microtype}          

\usepackage{mathtools}

\newcommand{\bolognaaff}{Department of Physics and Astronomy, University of Bologna and INFN-Bologna, 40126 Bologna, Italy}
\newcommand{\chicagoaff}{Department of Physics \& Kavli Institute for Cosmological Physics, University of Chicago, Chicago, IL 60637, USA}
\newcommand{\coimbraaff}{LIBPhys, Department of Physics, University of Coimbra, 3004-516 Coimbra, Portugal}
\newcommand{\columbiaaff}{Physics Department, Columbia University, New York, NY 10027, USA}
\newcommand{\lngsaff}{INFN-Laboratori Nazionali del Gran Sasso and Gran Sasso Science Institute, 67100 L'Aquila, Italy}
\newcommand{\mainzaff}{Institut f\"ur Physik \& Exzellenzcluster PRISMA$^{+}$, Johannes Gutenberg-Universit\"at Mainz, 55099 Mainz, Germany}
\newcommand{\heidelbergaff}{Max-Planck-Institut f\"ur Kernphysik, 69117 Heidelberg, Germany}
\newcommand{\munsteraff}{Institut f\"ur Kernphysik, Westf\"alische Wilhelms-Universit\"at M\"unster, 48149 M\"unster, Germany}
\newcommand{\nikhefaff}{Nikhef and the University of Amsterdam, Science Park, 1098XG Amsterdam, Netherlands}
\newcommand{\nyuadaff}{New York University Abu Dhabi - Center for Astro, Particle and Planetary Physics, Abu Dhabi, United Arab Emirates}
\newcommand{\purdueaff}{Department of Physics and Astronomy, Purdue University, West Lafayette, IN 47907, USA}
\newcommand{\riceaff}{Department of Physics and Astronomy, Rice University, Houston, TX 77005, USA}
\newcommand{\stockholmaff}{Oskar Klein Centre, Department of Physics, Stockholm University, AlbaNova, Stockholm SE-10691, Sweden}
\newcommand{\subatechaff}{SUBATECH, IMT Atlantique, CNRS/IN2P3, Nantes Universit\'e, Nantes 44307, France}
\newcommand{\torinoaff}{INAF-Astrophysical Observatory of Torino, Department of Physics, University  of  Torino and  INFN-Torino,  10125  Torino,  Italy}
\newcommand{\ucsdaff}{Department of Physics, University of California San Diego, La Jolla, CA 92093, USA}
\newcommand{\wisaff}{Department of Particle Physics and Astrophysics, Weizmann Institute of Science, Rehovot 7610001, Israel}
\newcommand{\zurichaff}{Physik-Institut, University of Z\"urich, 8057  Z\"urich, Switzerland}
\newcommand{\parisaff}{LPNHE, Sorbonne Universit\'{e}, CNRS/IN2P3, 75005 Paris, France}
\newcommand{\freiburgaff}{Physikalisches Institut, Universit\"at Freiburg, 79104 Freiburg, Germany}
\newcommand{\napelsaff}{Department of Physics ``Ettore Pancini'', University of Napoli and INFN-Napoli, 80126 Napoli, Italy}
\newcommand{\nagoyaaff}{Kobayashi-Maskawa Institute for the Origin of Particles and the Universe, and Institute for Space-Earth Environmental Research, Nagoya University, Furo-cho, Chikusa-ku, Nagoya, Aichi 464-8602, Japan}
\newcommand{\laquilaaff}{Department of Physics and Chemistry, University of L'Aquila, 67100 L'Aquila, Italy}
\newcommand{\tokyoaff}{Kamioka Observatory, Institute for Cosmic Ray Research, and Kavli Institute for the Physics and Mathematics of the Universe (WPI), University of Tokyo, Higashi-Mozumi, Kamioka, Hida, Gifu 506-1205, Japan}
\newcommand{\kobeaff}{Department of Physics, Kobe University, Kobe, Hyogo 657-8501, Japan}

\newcommand{\kitaff}{Institute for Astroparticle Physics, Karlsruhe Institute of Technology, 76021 Karlsruhe, Germany}
\newcommand{\tsinghuaaff}{Department of Physics \& Center for High Energy Physics, Tsinghua University, Beijing 100084, China}
\newcommand{\alsoatferraraaff}{INFN, Sez. di Ferrara and Dip. di Fisica e Scienze della Terra, Universit\`a di Ferrara, via G. Saragat 1, Edificio C, I-44122 Ferrara (FE), Italy}
\newcommand{\alsoatcoimbrapoliaff}{Coimbra Polytechnic - ISEC, 3030-199 Coimbra, Portugal}
\newcommand{\alsoatuniheidelbergaff}{Physikalisches Institut, Universit\"at Heidelberg, Heidelberg, Germany}

\begin{document}

\title{Effective Field Theory and Inelastic Dark Matter
Results from XENON1T}

\author{E.~Aprile}\affiliation{\columbiaaff}
\author{K.~Abe}\affiliation{\tokyoaff}
\author{F.~Agostini}\affiliation{\bolognaaff}
\author{S.~Ahmed Maouloud}\affiliation{\parisaff}
\author{L.~Althueser}\affiliation{\munsteraff}
\author{B.~Andrieu}\affiliation{\parisaff}
\author{E.~Angelino}\affiliation{\torinoaff}
\author{J.~R.~Angevaare}\affiliation{\nikhefaff}
\author{V.~C.~Antochi}\email[]{cristian.antochi@fysik.su.se}\affiliation{\stockholmaff}
\author{D.~Ant\'on Martin}\affiliation{\chicagoaff}
% ...

\collaboration{XENON Collaboration}
\email[]{xenon@lngs.infn.it}
\noaffiliation

\begin{abstract}
    In this work, we expand on the XENON1T nuclear recoil searches to study the individual signals of dark matter interactions from operators up to dimension-eight in a Chiral Effective Field Theory (ChEFT) and a model of inelastic dark matter (iDM). We analyze data from two science runs of the XENON1T detector totaling 1\,tonne$\times$year exposure. For these analyses, we extended the region of interest from [4.9, 40.9]$\,$keV$_{\text{NR}}$ to [4.9, 54.4]$\,$keV$_{\text{NR}}$ to enhance our sensitivity for signals that peak at nonzero energies. We show that the data is consistent with the background-only hypothesis, with a small background over-fluctuation observed peaking between 20 and 50$\,$keV$_{\text{NR}}$, resulting in a maximum local discovery significance of 1.7\,$\sigma$ for the Vector$\otimes$Vector$_{\text{strange}}$ ($VV_s$) ChEFT channel for a dark matter particle of 70$\,$GeV/c$^2$, and $1.8\,\sigma$ for an iDM particle of 50$\,$GeV/c$^2$ with a mass splitting of 100$\,$keV/c$^2$. For each model, we report 90\,\% confidence level (CL) upper limits. We also report upper limits on three benchmark models of dark matter interaction using ChEFT where we investigate the effect of isospin-breaking interactions. We observe rate-driven cancellations in regions of the isospin-breaking couplings, leading to up to 6 orders of magnitude weaker upper limits with respect to the isospin-conserving case.
\end{abstract}
\maketitle

\section{Introduction}
Astrophysical and cosmological observations indicate that roughly 80\% of the matter in the universe is dark matter (DM) \cite{Planck:2018vyg}, the nature of which is still unknown. The most promising particle DM hypothesis is that of the Weakly Interacting Massive Particle (WIMP)\cite{bertone2005particle,roszkowski2018wimp}. A number of experiments have been proposed and built for direct detection of WIMPs, among them are liquid xenon (LXe) dual-phase Time Projection Chamber (TPC) experiments such as XENON1T \cite{aprile2017xenon1t}, LUX \cite{mckinsey2010lux}, PandaX-II \cite{tan2016dark}. These experiments have probed WIMP masses above 6$\,$GeV/c$^2$ and put constraints on the WIMP-nucleus cross section for Spin-Independent (SI) and Spin-Dependent (SD) interactions \cite{collaboration2018dark,collaboration2019constraining,akerib2017results,akerib2017limits,cui2017dark,xia2019pandax}. 
New multi-tonne dual-phase TPC direct detection experiments such as XENONnT \cite{aprile2020projected}, LZ \cite{akerib2020projected,LZ:2022ufs} and PandaX-4T \cite{meng2021dark} are currently operating, LZ and PandaX-4T having already provided their first results, further probing the WIMP hypothesis. \par
The SI and SD analyses make some simplified assumptions about the interactions, considering them at the leading-order only, either with no spin dependence or coupling to the total nuclear spin \cite{Klos:2013rwa}, and a simple WIMP model with no internal degrees of freedom. In this work we expand the search to consider the contributions of individual operators of a Chiral Effective Field Theory (ChEFT) framework \cite{hoferichter2015chiral,hoferichter2016analysis,hoferichter2019nuclear,bishara2017chiral,bishara2017quarks} and report upper limits on the fundamental Wilson coefficients and on the physics scale arising from various coupling channels. Furthermore, we study the case of inelastic dark matter (iDM), where the DM particle has non-negligible internal degrees of freedom, when scattering off xenon nuclei within the XENON1T detector.

\subsection{Chiral Effective Field Theory Frameworks}
\begin{table*}
    \centering
    \begin{tabular}{c|c|c|c|c|c}
    \hline\hline
         Type & Abbrev. & Operator  & Dimension & Coherent & Coefficients \\
          & &($\mathcal{Q}$) & & enhancement & \\
         \hline
         Magnetic Dipole & - & $\Bar{\chi}\sigma^{\mu\nu}\chi F_{\mu\nu}$ & 5 & Partial & $C_F$\\
         Electric Dipole & - & $\Bar{\chi}\sigma^{\mu\nu}\chi \Tilde{F}_{\mu\nu}$ & 5 & Yes & $\tilde{C}_F$\\
         Vector$\otimes$Vector & $VV$ & $\Bar{\chi}\gamma^\mu\chi\Bar{q}\gamma_\mu q$ & 6 & Yes & $C_{u,d,s}^{VV}$\\
         Axial-vector$\otimes$Vector & $AV$ & $\Bar{\chi} \gamma^\mu \gamma_5 \chi\Bar{q}\gamma_\mu q$& 6 & Yes & $C_{u,d}^{AV}$\\
         Tensor$\otimes$Tensor & $TT$ & $\Bar{\chi}\sigma^{\mu\nu}\chi \Bar{q}\sigma_{\mu\nu}q$& 6 & Yes & $C_{u,d,s}^{TT}$\\
         Pseudo-tensor$\otimes$Tensor& $\widetilde{TT}$ & $\Bar{\chi}\sigma^{\mu\nu}i\gamma_5 \chi \Bar{q}\sigma_{\mu\nu}q$& 6& Yes& $\tilde{C}_{u,d,s}^{TT}$\\
         Scalar$\otimes$Scalar & $SS$& $\Bar{\chi}\chi m_q \Bar{q}q$ & 7& Yes& $C_{u,d,s}^{SS}$\\
         Scalar-gluon & $S_g$ & $\alpha_s\Bar{\chi}\chi G^a_{\mu\nu}G_a^{\mu\nu}$& 7& Yes& $C_g^{S}$\\
         Pseudo-scalar - gluon & $\tilde{S}_g$ &$\alpha_s\Bar{\chi}i\gamma_5 \chi G^a_{\mu\nu}G_a^{\mu\nu}$& 7& Yes & $\tilde{C}_g^{S}$\\
         Pseudo-scalar$\otimes$Scalar & $PS$& $ \Bar{\chi}i\gamma_5\chi m_q\Bar{q}q$ & 7 & Yes &$C_{u,d,s}^{PS}$\\
         Spin-2& - & $\Bar{\chi}\gamma_\mu i \partial_\nu\chi\Bar{\theta}^{\mu\nu}_{q(g)}$ & 8 & Yes &$C_{u,d,s,g}^{(2)}$\\
         \hline
         Axial-vector$\otimes$Axial-vector & $AA$ &  $\Bar{\chi} \gamma^\mu \gamma_5 \chi\Bar{q}\gamma_\mu \gamma_5 q$& 6 & No & $C_{u,d,s}^{AA}$\\
         \hline \hline
    \end{tabular}
    \caption{Summary of the ChEFT operators considered in the ChEFT analysis of this work, showing the abbreviation used in the paper, the analytical expression of the operators, the dimension, and the respective coefficients. The $AA$ operator is shown here, but it is not used for the single operator analysis, since it does not lead to a coherent enhancement in the nuclear response.}
    \label{tab:operators}
\end{table*}

Recent direct detection DM Effective Field Theory (EFT) analyses were performed within the non-relativistic EFT (NREFT) framework \cite{aprile2017effective,akerib2021constraints,schneck2015dark} described in \cite{Fitzpatrick_2013}, which constructs the effective Lagrangian of the WIMP-nucleon interaction considering the non-relativistic Galilean-invariant operators. While such analyses are useful to constrain nuclear responses beyond the standard SI and SD models, they are difficult to interpret in terms of fundamental interactions. A chiral analysis can be mapped onto NREFT operators, at the single-nucleon level. The relations between ChEFT and NREFT operators, however, show that the NREFT operators are not independent due to QCD effects \cite{hoferichter2015chiral}.\\
ChEFT expands the Quantum Chromo-dynamics (QCD) Lagrangian in orders of the momentum exchange over the chiral symmetry breaking QCD scale. The obtained ordering preserves QCD symmetries at low energies and captures the importance of pions and two-body currents \cite{hoferichter2016analysis,hoferichter2019nuclear}. The chiral regime is well justified to study WIMP-nucleus interactions, given that momentum transfer is typically of the same order of the pion mass, a scale relevant for momenta in heavy nuclei such as xenon \cite{Aalbers:2022dzr}.\\
From the particle physics perspective, low energy DM interactions can be effectively parameterized in terms of the lightest three quark flavors, $up$ ($u$), $down$ ($d$) and $strange$ ($s$), the gluon and the photon. The interactions are ordered according to the dimension of the operators and the effective Lagrangian is written in the form of
\begin{equation}
    \mathcal{L}_{\chi\:EFT}=\sum_{\text{d},a,q(g)}\frac{C^{a,(\text{d})}_{q(g)}}{\Lambda^{\text{d}-4}}\mathcal{Q}^{(\text{d})}_{a,q(g)},
    \label{eq:gen_lagrangian}
\end{equation}
where the $C^{a,(\text{d})}_{q(g)}$ are dimensionless Wilson coefficients, $\Lambda$ is the physics scale of the interaction and $\mathcal{Q}^{(\text{d})}_{a,q(g)}$ the assosiated operator of dimension $\text{d}$. The sum in Eq.\,\ref{eq:gen_lagrangian} runs over the different operator type, $a$, the quarks (gluons), $q$ ($g$), and the operator dimensions, $\text{d}$. In this scheme, the Wilson coefficients are constants that contain all the information about the interaction. For a model independent analysis they can be freely varied. By going through a series of matchings, constraints on low energy ChEFT operators can be extended to high-energy interactions, easing the comparison with accelerator constraints \cite{Crivellin:2014qxa,DEramo:2014nmf}.\\
There are two different approaches in performing a ChEFT analysis: one is starting from the nuclear level perspective, taking the SI cross section and reconstructing the nuclear response from chiral level, focusing on the chiral contributions to the nuclear structure factors \cite{hoferichter2015chiral,hoferichter2016analysis,hoferichter2019nuclear}, the other is finding a complete basis of ChEFT operators in the three quark flavor EFT and create a matching to the non-relativistic single nucleon EFT level \cite{Bishara:2016hek,Bishara:2017nnn,Bishara:2017pfq,Bishara:2018vix,bishara2020renormalization}. These two approaches have lead to the development of two complementary frameworks, respectively, the Generalised SI ChEFT framework \cite{hoferichter2019nuclear} and the DirectDM framework \cite{Bishara:2017nnn}.
In this work we perform a ChEFT analysis of all the chiral operators that contribute to the nuclear response in a coherent way, using the full information about the nuclear form factors from \cite{hoferichter2019nuclear}. We consider operators up to dimension-eight, coupled to a large-scale shell model computation \cite{Caurier:2007wq,Menendez:2008jp,Menendez:2012tm,Vietze:2014vsa,Caurier:2004gf} of the nuclear structure factors to compute possible WIMP-nucleus interactions observable in the XENON1T detector, and set constraints on the Wilson coefficients and the interaction scale, $\Lambda$. In Tab.\,\ref{tab:operators} we show a list of the operators, the terminology and the coefficients we investigate in this work.\\
For a set of operators that appear at leading order in the most common WIMP models, we present constraints obtained with both the Generalised SI framework and the DirectDM framework. The full list of operators, the matching and differences between the two frameworks are detailed in Appendix\; \ref{app:GenSI-DirDM}.

\subsubsection*{Isospin-breaking Couplings}
Besides the constraints on the individual Wilson coefficients, we include the study of three benchmark models of WIMP interactions corresponding to the most popular DM models, where the leading contributions arise from a single type of couplings within ChEFT: 
\begin{itemize}
    \item vector mediator for Majorana DM, with leading contribution from the $AV$ operators \cite{matsumoto2014singlet},
    \item vector mediator for Dirac DM, with leading contribution from the $VV$ operators,
    \item and scalar mediator for fermion DM, with leading contribution from the $SS$ operators.
\end{itemize}
In these models we study the effect of isospin-breaking interactions by changing the value of the $u$ and $d$ Wilson coefficients and computing the limits for various combinations of the two, neglecting possible contributions from $s$ and $g$ couplings.\\
Turning on both $u$ and $d$ coefficients, for a given ratio $r=C^{a}_u/C^{a}_d$, we can set constraints on one of the coefficients, which can then be extrapolated in constraints on the $C^{a}_u$, $C^{a}_d$ plane, given the symmetry under parity transformation.\\

In the treatment of the vector mediator for Majorana DM model, due to operators above the weak scale matching onto both $AV$ and Axial-vector$\otimes$Axial-vector ($AA$) operators, the $AA$ contribution cannot be set to zero. Thus, to retain the freedom to vary $C^{AV}_d$ and $C^{AV}_u$ independently, we set $C^{AA}_u=0$ and $C^{AA}_d = C^{AV}_d - C^{AV}_u$, to maintain the relations of the above-weak-scale operators \cite{bishara2020renormalization}, and study the limit on the signal rate as a function of the ratio $C^{AV}_u/C^{AV}_d$. A more detailed description of the treatment of the vector mediator for Majorana DM model can be found in Appendix\;\ref{ap:Majorana}.

\begin{figure*}
    
    \includegraphics[width=6.75in]{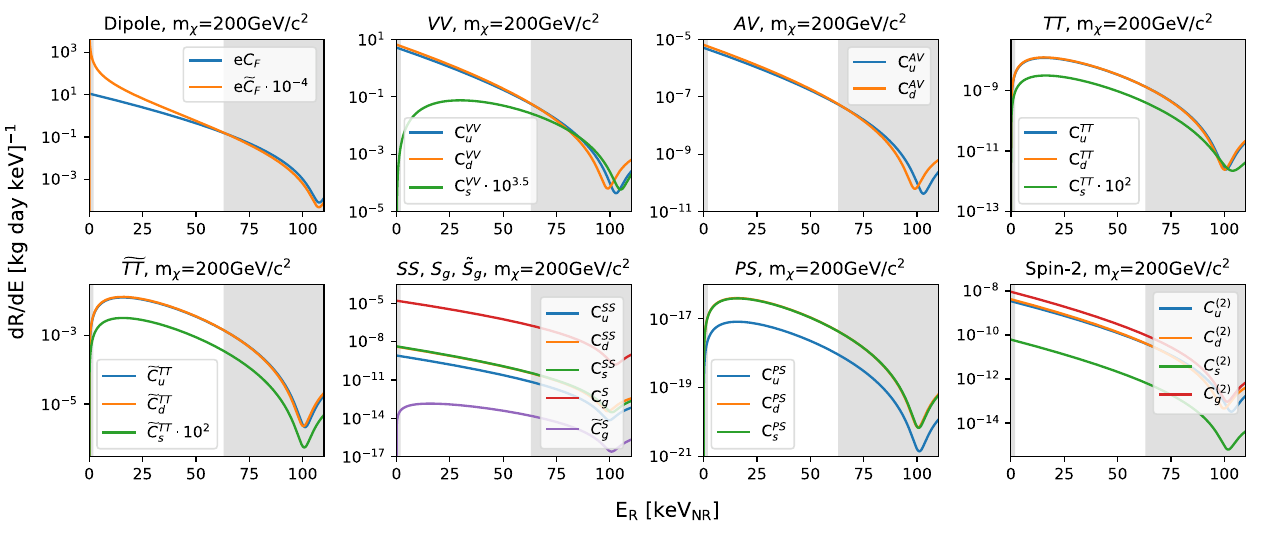}
    \caption{Differential rate spectra for the ChEFT operators investigated in this work, divided according to the interaction type. They are obtained by setting each Wilson coefficient to 1 and $\Lambda = 1\,$TeV. The shaded region marks the energy range where the signal acceptance is below 10\%. For some spectra, the coefficients were multiplied by a factor written in the legend for plotting purposes.}
    \label{fig:diff_rate_cheft}
\end{figure*}
\begin{figure*}
    \centering
    \includegraphics[width=0.8\textwidth, center]{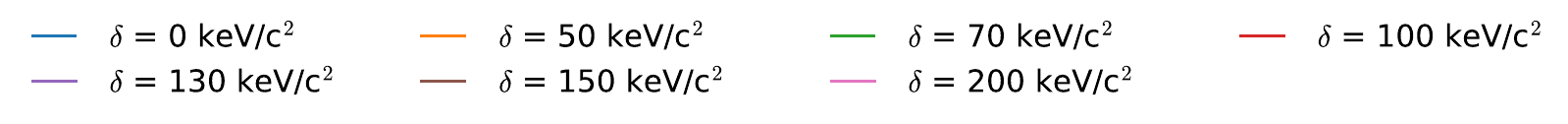}
    \includegraphics[width=6.75in]{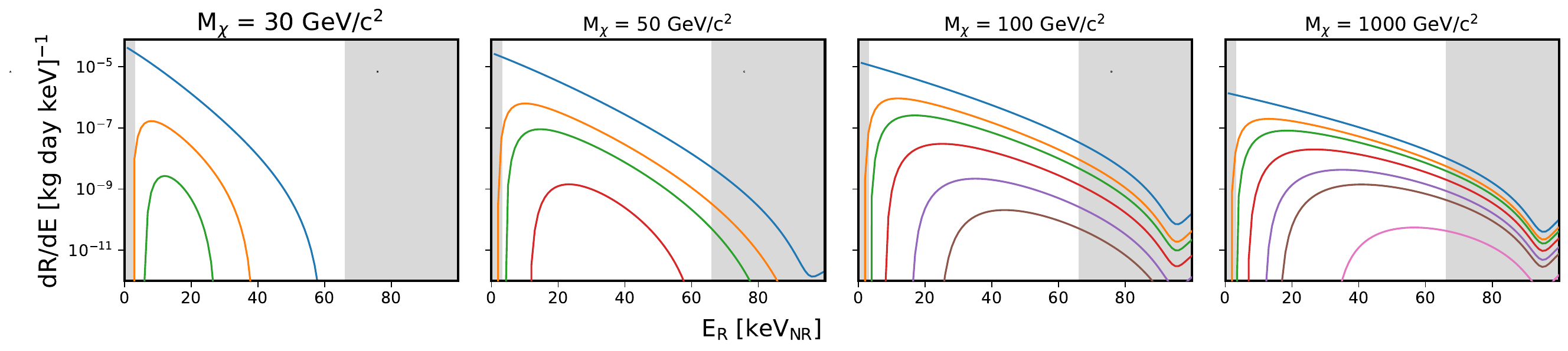}
    \caption{Differential rate spectra for a few representative iDM models investigated in this work. The grey area indicates the region where the detection efficiency is below 10\%. The parameter space explored was selected to include models peaked within the selected energy region. }
    \label{fig:diff_rate_idm}
\end{figure*}
\subsection{WIMPs with Structure - Inelastic Dark Matter}
A common assumption used in WIMP searches is that the DM internal degrees of freedom either dont exist or are not relavent on the energy scale being probed. This simplifies the analysis greatly, but many alternative models, in which this assumption has been relaxed, have been proposed, e.g. \cite{cui_candidates_2009}. A primary motivation to study the signature of these models in DM experiments was the tension between the measured DAMA \cite{bernabei_dama/libra_2008} spectrum and results from other experiments \cite{Tucker-Smith:2001myb,Lang:2010cd}. Although recent results \cite{Amare:2021yyu,COSINE-100:2021xqn} favor a non-DM explanation to the DAMA result, iDM models remain are interesting in their own right due to their unique spectra. In these models, the DM particle has an excited state which it transitions to during scattering off Standard Model (SM) nuclei, $ \chi N \to \chi^* N $, while elastic scattering is forbidden or highly suppressed. The mass splitting between the DM particle states $\chi$ and $\chi^*$ is usually denoted as $\delta$ and introduces a minimum velocity for the WIMPs to scatter in our detector and deposit an energy $E_{\text{R}}$ 
\begin{equation}
    \beta_{min} = \sqrt{\frac{1}{2M_\text{N} E_{\text{R}}}} \left(\frac{M_\text{N} E_{\text{R}}}{\mu} + \delta\right).
    \label{eq:idm_vmin}
\end{equation}
This minimum velocity can significantly suppress the expected number of events at low recoil-energies, resulting in a recoil spectrum peaked at nonzero E$_{\text{R}}$. In the limit of very small $\delta$, the expected recoil spectrum reproduces the standard WIMP.

\subsection{Recoil Spectra}

The recoil spectra for the ChEFT analysis are obtained by setting each single Wilson coefficient to 1 and all the others to 0, and the reference physics scale to $\Lambda_{\text{ref}}=1\,$TeV in Eq.\,\ref{eq:gen_lagrangian}. 
Turning on one Wilson coefficient at a time makes it possible to study the direct contribution of the fundamental interaction to the nuclear response, since the differential rate,
\begin{equation}
    \frac{dR}{dE_\text{R}}=\frac{\rho}{2\pi m_{\chi}}\times|\mathcal{F}(q^2)\big|^2\times\int_{v_\text{min}(E_\text{R})}^{\infty}
\frac{f({\bf v})}{v}d^3v,
\label{eq:rate}
\end{equation}
contains the momentum transfer ($q^2$) dependent nuclear response $|\mathcal{F}(q^2)|^2$, which is directly proportional to the single activated Wilson coefficient as $ |\mathcal{F}(q^2)|^2 \propto |C^i/\Lambda^{d-4}|^2$.\\
To perform the rate computation in Eq.\,\ref{eq:rate}, we consider the DM velocity distribution, $f(\bf{v})$, to be described by a Maxwell distribution, truncated at the DM escape velocity, $v_\text{esc}=544\,$km/s, and employ the standard halo model (SHM) quantities of DM density $\rho= 0.3\,\text{GeV}/(\text{c}^2\times \text{cm}^3)$,  $v_0 =220\,$km/s, and Earth velocity of $v_\text{E}=232\,$km/s,
and use the tools provided in \cite{hoferichter2019nuclear} to compute the nuclear response. Fig.\,\ref{fig:diff_rate_cheft} shows the differential rate spectra obtained for all the investigated channels of interaction within the Generalised SI ChEFT framework. For the set of operators investigated with both ChEFT frameworks, we computed the spectra using both the Generalised SI ChEFT framework and the DirectDM package \cite{Bishara:2017nnn} and the aid of the nuclear response from \cite{Anand_2014} using the DMFormFactor package. \\
The recoil spectra for iDM on xenon were calculated using a Mathematica package based on DMFormFactor \cite{Fitzpatrick_2013,Anand_2014} modified by \cite{barello2014} to use operator $\mathcal{O}_1$ of the low energy NREFT, corresponding to the SI interaction, and to impose the threshold in Eq. \ref{eq:idm_vmin} on the energy transfer. The recoil spectra were calculated for the xenon isotopes in the detector ($^{128}$Xe, $^{129}$Xe, $^{130}$Xe, $^{131}$Xe, $^{132}$Xe, $^{134}$Xe and $^{136}$Xe) and weighted by their relative abundance. 
The differential recoil spectra for a selection of parameter space the detector is sensitive to are shown in Fig. \ref{fig:diff_rate_idm}.

\section{XENON1T}
The XENON1T detector was a direct DM detection experiment that consisted of a dual-phase xenon TPC, with height of 97\,cm and diameter of 95.8\,cm \cite{aprile2017xenon1t}. The detector was placed at the underground facility of Laboratori Nazionali del Gran Sasso (LNGS) in Italy, under the Gran Sasso massif, which provides rock shielding of approximately 3600\,m of water equivalent. The TPC was surrounded by a $\sim700\,$t water Cherenkov detector operated as a muon-veto \cite{XENON1T:2014eqx} that also provided shielding from environmental radiation. Interactions within the instrumented volume can produce photons, ionization electrons and heat. The detector was designed to collect both electrons and photons using two photomultiplier tube (PMT) arrays located on the top and bottom of the detector \cite{Barrow:2016doe}. The photons emitted from the interaction were measured directly using the PMTs, while the electrons were drifted up to the gaseous phase of the TPC via a drift field of $\sim$100 V/cm (120 $\pm$ 8 V/cm in SR0 and 81 $\pm$ 6 V/cm in SR1). The drifted electrons are then accelerated through the gas phase in a $\sim$10 kV/cm field to produce proportional scintillation light, which was then measured by the same PMT arrays. The prompt signal generated by primary photons is commonly referred to as S1 and the delayed signal produced by the extracted electrons is denoted S2. Interactions of different types produce different relative sizes of S2 and S1 signals in the detector, allowing to use the ratio of the signal amplitudes, S2/S1, for  discrimination between Nuclear Recoils (NR) and Electronic Recoils (ER). The NR-ER discrimination significantly increases the detector's sensitivity to NR signals like WIMPs. The spatial position of the interaction is reconstructed using the time difference between the signals for the Z dimension and a pattern of light detected on the top PMT array for the X and Y dimensions. Since the detector response was not spatially uniform, the measured values (S1 and S2) were corrected for these differences and then used in analysis which is performed in the corrected space cS1 and cS2. This spatial dependence of the detector response was calibrated by introducing radioactive sources ($^{83\text{m}}$Kr) into the xenon and observing the signal produced by the spatial dependence of the spatially uniform radioactive decays. In the analysis, we use the corrected S2 signal only from the bottom PMT array (cS2$_\text{b}$) which was found empirically to provide better energy reconstruction of the events.

\section{Data analysis}

This work re-analyzes the data from two science runs (SR0 and SR1) explored in the SI WIMP analysis \cite{collaboration2019xenon1t}, extending, however, the region of interest (ROI) of the analysis up to 100$\,$ photo-electrons (PE) in cS1, resulting in an extension of the energy range from $[4.9, 40.9]\,$keV$_\text{NR}$ to $[4.9, 54.4]\,$keV$_\text{NR}$, as shown in Fig.\,\ref{fig:acceptance_plot}. The combined data collected by the XENON1T detector over two science runs, SR0 and SR1, with livetimes of 32.1 days and 246.7 days respectively, and a fiducial volume, containing 1.3 tonnes of liquid xenon, provide an exposure of 1 tonne$\times$year.

\subsection{Analysis region}
\begin{figure}
    \includegraphics[width=.48\textwidth]{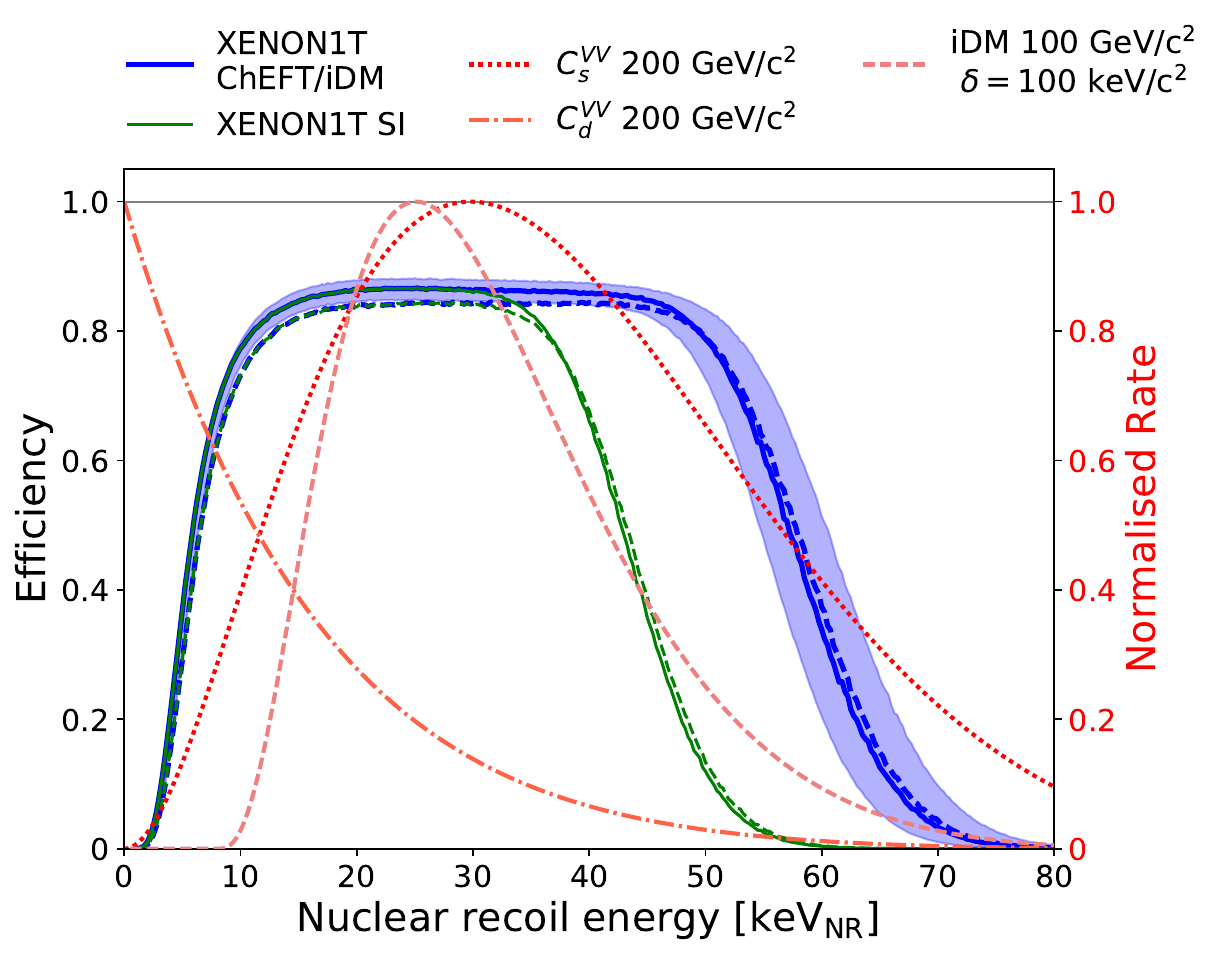}
    \caption{Median analysis efficiency versus nuclear recoil energy for the current analysis (blue) for SR0 (dashed blue line) and SR1 (solid blue line) with the $1\sigma$ uncertainty band (blue band) shown for SR1 only, SRO band being almost identical to it. The dashed and solid green lines show the median efficiency for the XENON1T SI analysis for respectively SR0 and SR1. In red we show spectra for the ChEFT $VV$ interactions for a $200\:$GeV/c$^2$ WIMP, $VV_d$ (dash-dotted line) and $VV_s$ (dotted line), and the iDM spectrum (dashed line) for a $100\:$GeV/c$^2$ WIMP with $\delta=100\:$keV/c$^2$. }
    \label{fig:acceptance_plot}
\end{figure}

The analysis was carried out in an increased energy region to increase the acceptance for signal models producing spectra that peak at finite recoil energies. The extension of the ROI resulted in a signal acceptance increase up to 20\,\% for the signals that peak at the non-zero energies, as in the case of the $TT$ and $VV_s$ operators for WIMP masses above 50$\,$GeV/c$^2$, and iDM models with a splitting up to 200\,keV/c$^2$. The total ROI range of the analysis was selected between 3 and 100$\,$PE in cS1, between 50 and 8000$\,$PE in cS2$_\text{b}$, and within a radius $R< 42.8\,$cm. These selections, together with data quality selections, result in a total detection efficiency shown in Fig. \ref{fig:acceptance_plot}.
An extension of the analysis region beyond $100\,$PE in cS1 could not be achieved due to the difficulty of correctly modelling both the signal and background distributions and not enough calibration statistics in this region.\\ 

The data in the region of 70-80$\,$PE in cS1 was previously used for the background model validation and thus was already fully unblinded, however the region between 80-100$\,$PE in cS1 maintained an NR band blinding cut prior to this analysis. This region was unblinded when the statistical framework was properly validated for this analysis. \\

\subsection{Calibration and Backgrounds}
The main expected background for this analysis consists of electronic recoils in the detector caused by background radiation, this is due to the tail of the ER band extending into the NR band, commonly referred to as ER leakage. The ER band was modeled according to the LXe emission model described in \cite{collaboration2019xenon1t}, and calibrated with radioactive sources diffused in the detector, such as $^{220}$Rn and $^{83\text{m}}$Kr \cite{aprile2019xenon1t}.\\
The spatial dependence of the background was modelled in the radial coordinate R, while the Z dependence was included by binning the data into two independent volumes, a low-background inner volume and the higher-background external volume.\\
The NR band was calibrated during dedicated calibration runs, using a deuterium-deuterium neutron generator and a $^{241}$AmBe neutron source.\\
The main NR background consists of neutrons produced in radioactive decays in the TPC materials, and was modelled using a full Geant4 \cite{GEANT4:2002zbu,Allison:2016lfl} simulation of the detector and detector materials, including a spatial radial dependence. For low energy NR recoils, coherently-enhanced neutrino-nucleus scattering (CE$\nu$NS) events are an irreducible background, and modeled considering the flux of $^{8}$B solar neutrinos.\\
ER events depositing energy close to the surfaces of the TPC, due to material gamma and beta radiation from contaminants in the $^{222}$Rn decay chain, can also produce detectable signals in the ROI. Charge loss along the TPC walls reduces the observed charge signal, bringing the events below the NR region in cS2$_\text{b}$. Such background will have a significant radial dependence and was modelled through a data driven adaptive kernel density estimation (KDE) model.\\
Finally, accidentally paired S1s and S2s (AC), were modelled by randomly pairing lone S1 and S2s.\\
In Table \ref{tab:my_label} we report the background-only best fit expectation value, obtained by fitting the data without the signal model, for the background sources in the extended analysis region.\\
We also investigated a-posteriori the effect of a possible additional mono-energetic ER background at $2.3\,$keV$_\text{ER}$ following the indication of a low energy ER excess observed in \cite{aprile2020excess}. The presence of such additional background does not affect the NR search considerably, with confidence interval results not changing beyond 5\,\% when this background component is added. Given the uncertain nature of this structure in the ER band and the limited effect that it would have on a NR search, this analysis does not include it as a new background.

\begin{figure*}
    \includegraphics[width=.9\textwidth]{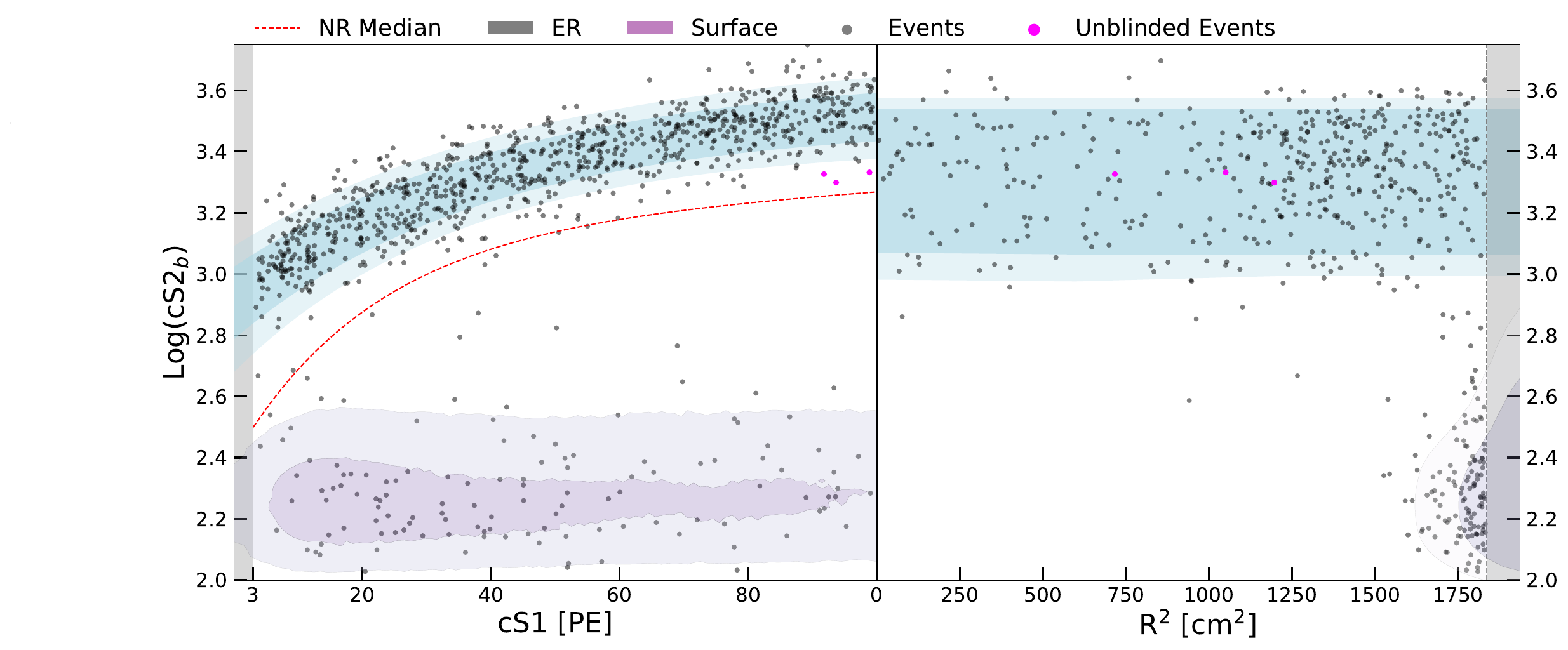}
    \caption{Event distribution in (cS1, cS2$_\text{b}$) (left) and (R$^2$, cS2$_\text{b}$) (right) parameter spaces. Projections of the two dominant backgrounds are shown as contours in light blue (ER) and purple (Surface), with darker and lighter shades that indicate the $1\,\sigma$ and $2\,\sigma$ contours respectively. The median of the nuclear recoil band is marked as a red line. The three magenta events in the NR band are included in this analysis as a result of expanding the ROI. The shaded regions mark parameter space excluded from the analysis.}
    \label{fig:signal_models_vs_event_distribution}
\end{figure*}

\begin{figure*}
    \includegraphics[width=.9\textwidth]{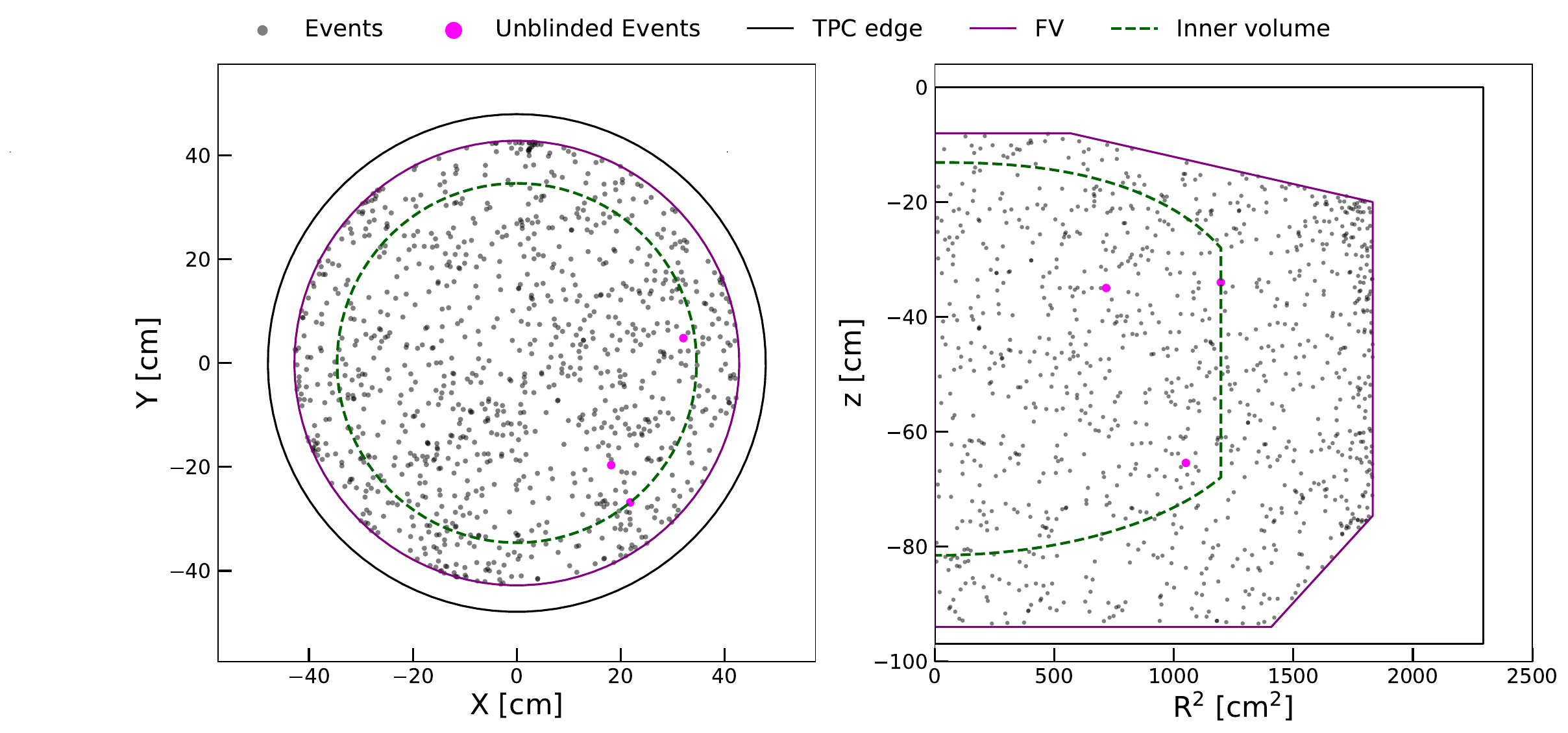}
    \caption{Spatial distribution of the detected events. Only events that pass all selection criteria and are within the fiducial mass are shown. Events that were unblinded in this analysis due the the expansion of the ROI are shown in magenta.
 The TPC
boundary (black line), 1.3\,t fiducial mass (purple) and 0.65\,t core
mass (green dashed) are shown.}
    \label{fig:event_spatial_distribution}
\end{figure*}

\begin{table}
    \centering
    \begin{tabular}{c c c}
    \hline\hline
       Background   & Best fit\\
       \hline
       ER  & $893\pm22$ \\
       neutron  & $1.55\pm0.71$ \\
       CE$\nu$NS &$0.054\pm0.015$ \\
       AC  & $0.51^{+0.28}_{-0.00}$ \\
       Surface & $133\pm12$\\
       \hline
       Total BG & $1028\pm25$\\
       \hline
       Data & $1032$\\
       \hline\hline
       
    \end{tabular}
    \caption{Background-only best fit expectation value of the total backgrounds for the combined SR0+SR1 datasets in the extended analysis region of [3,100]$\,$PE in cS1, and the total number of events observed in the SR0+SR1 data sets.}
    
    \label{tab:my_label}
\end{table}

\subsection{Inference}
The statistical analysis was performed in the three-dimensional space cS1, cS2$_\text{b}$ and R, as shown in Fig.\,\ref{fig:signal_models_vs_event_distribution}, where the background was modelled independently in the low-background inner volume and in the rest of the fiducial volume, shown in Fig.\,\ref{fig:event_spatial_distribution}. A single parameter-of-interest $\mu$ (the number of expected signal events) was inferred, where the DM particle mass were sampled between 10 GeV/c$^2$ and 10 TeV/c$^2$, calculating the local significance and limit for each and interpolating. The iDM parameter space was sampled in mass splitting from 0 to 200 keV/c$^2$. In the eventuality of a model having a local significance above the $3\,\sigma$ threshold, we would have reported also the global significance including the trial factor. \\
The limits and the discovery significances were computed using the profile log-likelihood ratio test statistic. As in the WIMP analysis \cite{collaboration2019xenon1t}, we used a combined unbinned likelihood for the two science runs, SR0 and SR1, with additional terms for the ER band calibration fit and ancillary measurements constraints for the background rate.
The profile-likelihood was constructed using the same null-hypothesis for all analyses, only replacing the signal model. The signal model itself was generated using a Markov Chain Monte Carlo (MCMC) simulation to model the expected detector response from the calculated spectra of events produced by the Generalised SI ChEFT  and DirectDM frameworks.\\
For this analysis we compute two-sided intervals based on the likelihood ratio test and in accordance to the previous XENON1T results, we set a $3\,\sigma$ discovery threshold above which we report both upper and lower limits. \\
A 15\,\% power constrained limit (PCL) was used to ensure limits are not set below the sensitivity of our detector, following the procedure described in \cite{cowan2011power}.\\
An additional safeguard nuisance parameter, as described in \cite{Priel:2016apy}, was added to the ER background model to prevent any potential mis-modelling in the signal region biasing the results.

\subsubsection*{Asymptotic cross checks}
\begin{figure}
    \includegraphics[width=3.375in]{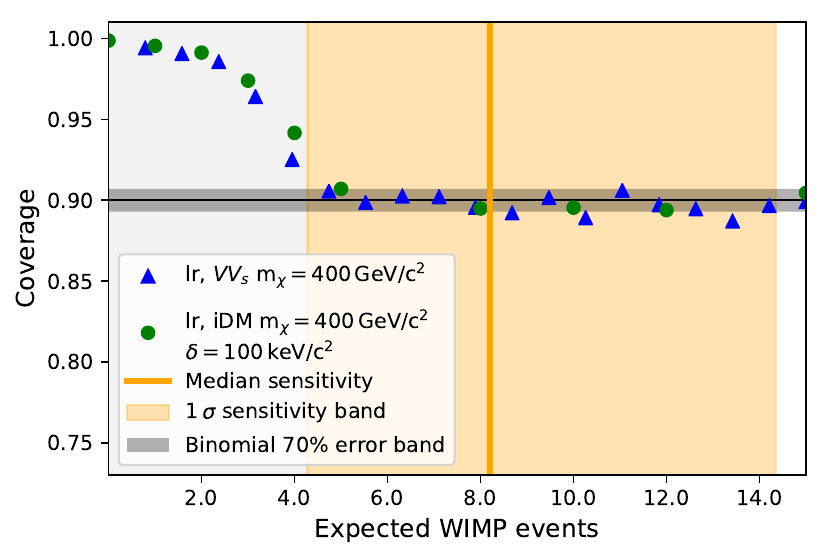}
    \caption{Coverage plot of the likelihood ratio (lr in legend) test statistic with $3\sigma$ threshold for reporting a two-sided interval for one example EFT operator ($VV_s$) and WIMP mass of $400\,$GeV/c$^2$ (blue triangles), and 400$\,$GeV/c$^2$ iDM with a splitting of $\delta=100\,$keV/c$^2$ (green dots). The median sensitivity and its 1$\sigma$ uncertainty are shown as orange solid line and orange shaded area, respectively. The black solid line indicates the $90\,\%$ coverage with the $70\,\%$ binomial error band (horizontal grey band). The vertical light-grey shaded area indicates the region where the power threshold is applied.}
    \label{fig:coverage}
\end{figure}

Given the large number of signal models, we compute confidence intervals and discovery significances assuming the test statistic follows the asymptotic distributions listed in \cite{Cowan:2010js} and for each model we perform coverage checks to verify that no significant undercoverage affects the result. The distribution of the discovery test statistic under the null hypothesis and the coverage of the confidence intervals were studied with toy Monte Carlo (MC) simulations. These studies have shown that for most signal models with a WIMP mass above 10$\,$GeV/c$^2$, there is no significant under-coverage in the sensitivity region, with slight over-coverage for signal expectation values of $\mu<4$. In Fig. \ref{fig:coverage} we show a coverage plot for the $VV_s$ signal model for a WIMP mass of 400$\,$GeV/c$^2$, and an iDM signal model for a particle of 400$\,$GeV/c$^2$ with a splitting of $\delta=200\,$keV/c$^2$, based on 2000 toy MCs each that show the typical coverage for different expectation values. \\

\section{Results}

\begin{figure*}
\centering
    \includegraphics[width=3.375in]{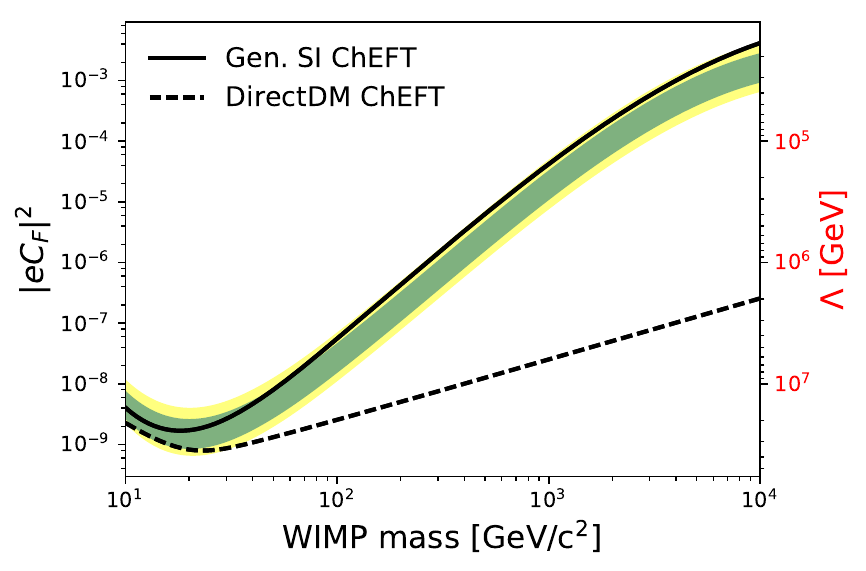}
    \includegraphics[width=3.375in]{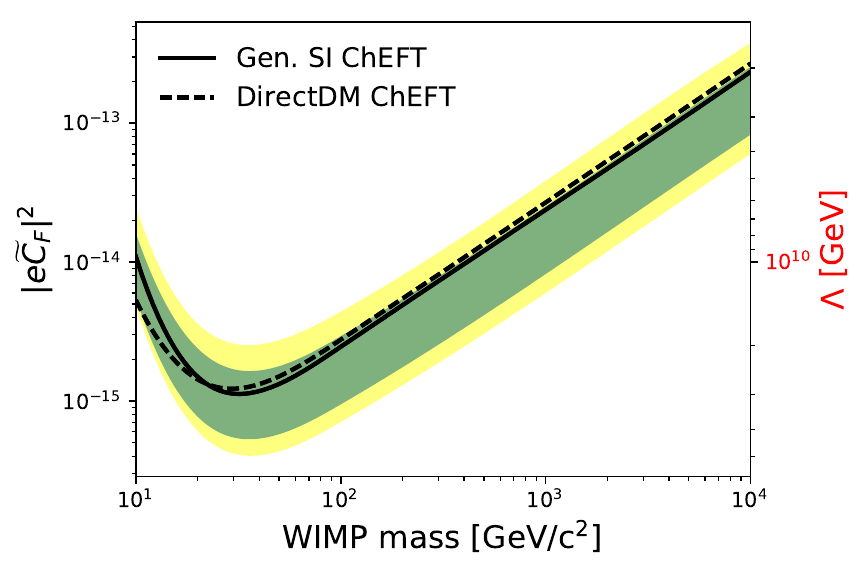}
    \caption{90\,\% CL upper limits (solid black lines) on the Wilson coefficients with physics scale set to 1 TeV (left scale) and the physics scale $\Lambda$ with the Wilson coefficient set to 1 (right, inverted scale, in red) for the magnetic (left panel) and electric (right panel) dipole couplings in the Generalised SI ChEFT framework, with the $1\,\sigma$ (green) and $2\,\sigma$ (yellow) sensitivity bands. For comparison we show the harmonized 90\,\% confidence level upper limits (black dashed lines) obtained with the DirectDM framework.}
    \label{fig:limits_EFT_dim5}
\end{figure*}
\begin{figure*}
\centering
    \includegraphics[width=3.375in]{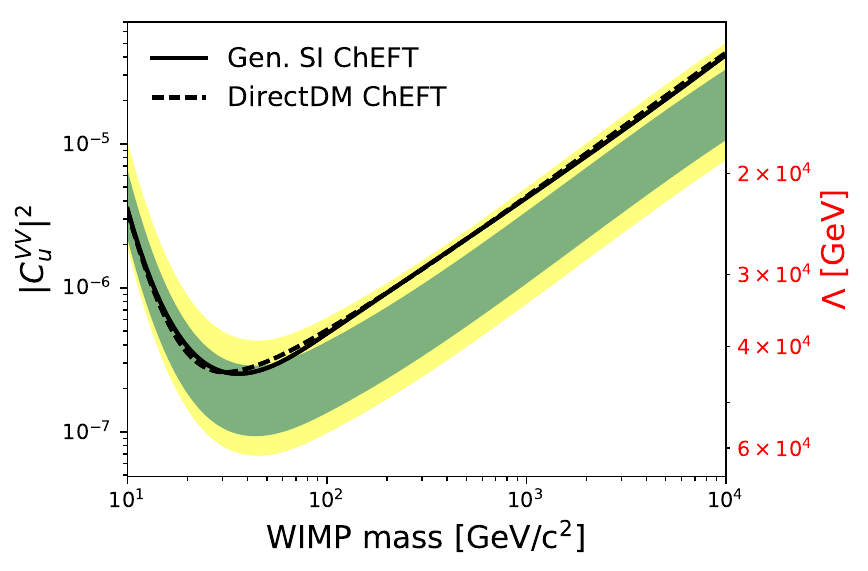}
    \includegraphics[width=3.375in]{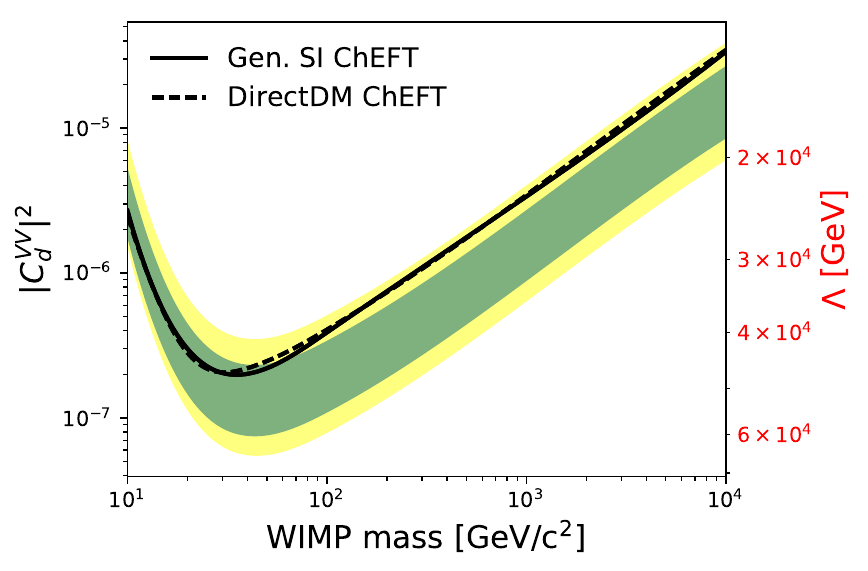}\\
    \includegraphics[width=3.375in]{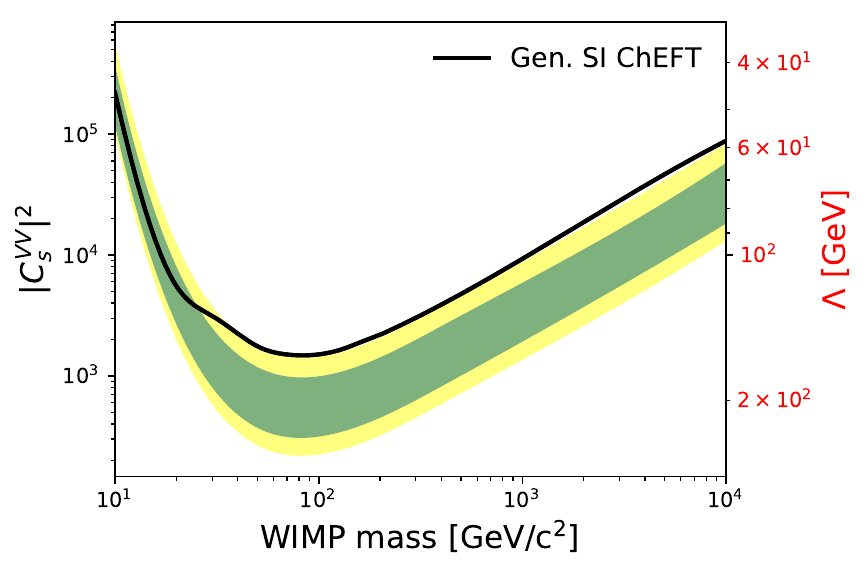}
    \includegraphics[width=3.375in]{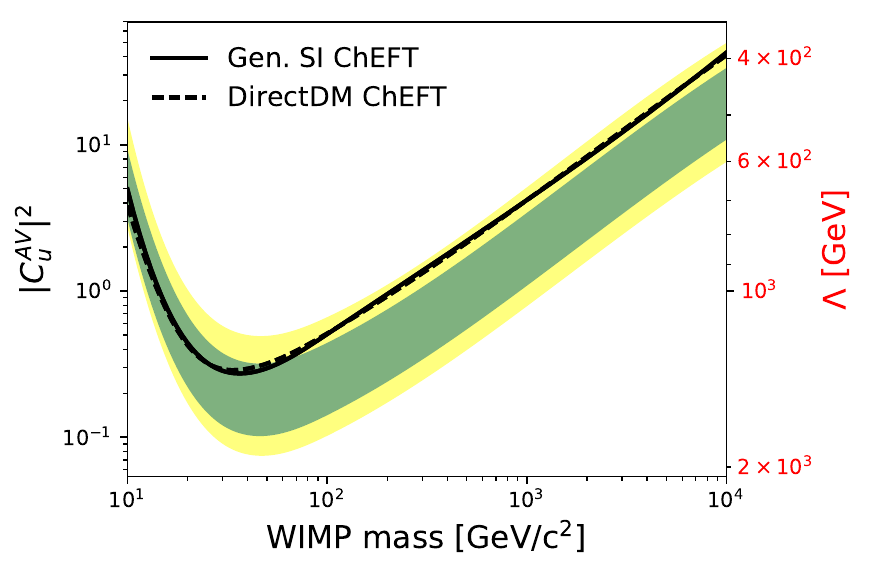}\\
    \includegraphics[width=3.375in]{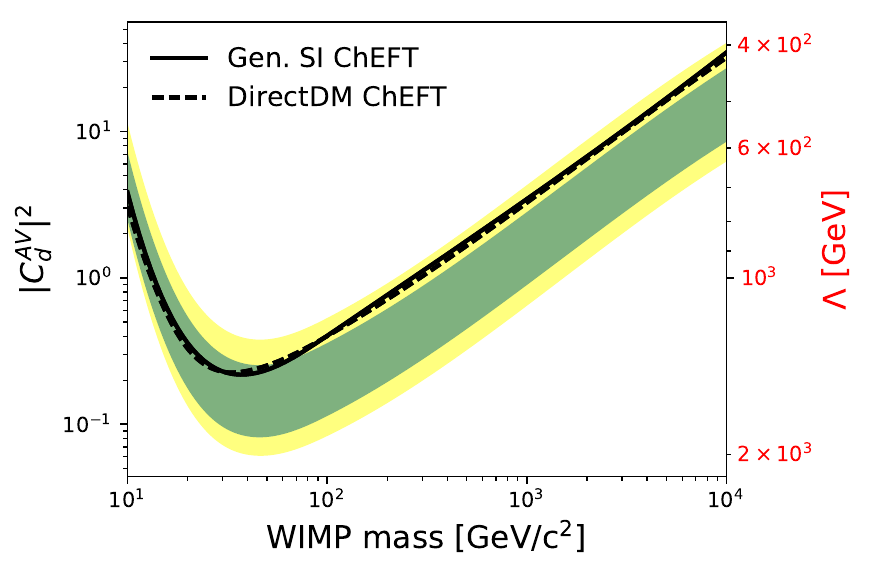}
    \caption{90\,\% CL upper limits (solid black lines) on the Wilson coefficients (left scale) and the physics scale $\Lambda$ (right, inverted scale, in red) for the $VV_u$ (top left), $VV_d$ (top right), $VV_s$ (centre left), $AV_u$ (centre right) and $AV_d$ (bottom) couplings in the Generalised SI ChEFT framework, with the $1\,\sigma$ (green) and $2\,\sigma$ (yellow) sensitivity bands. For comparison we show the harmonized 90\,\% confidence level upper limits (black dashed lines) obtained with the DirectDM framework for the $VV_u$, $VV_d$, $AV_u$ and $AV_d$ operators.}
    \label{fig:limits_EFT_dim6_1}
\end{figure*}
\begin{figure*}
\centering
    \includegraphics[width=3.375in]{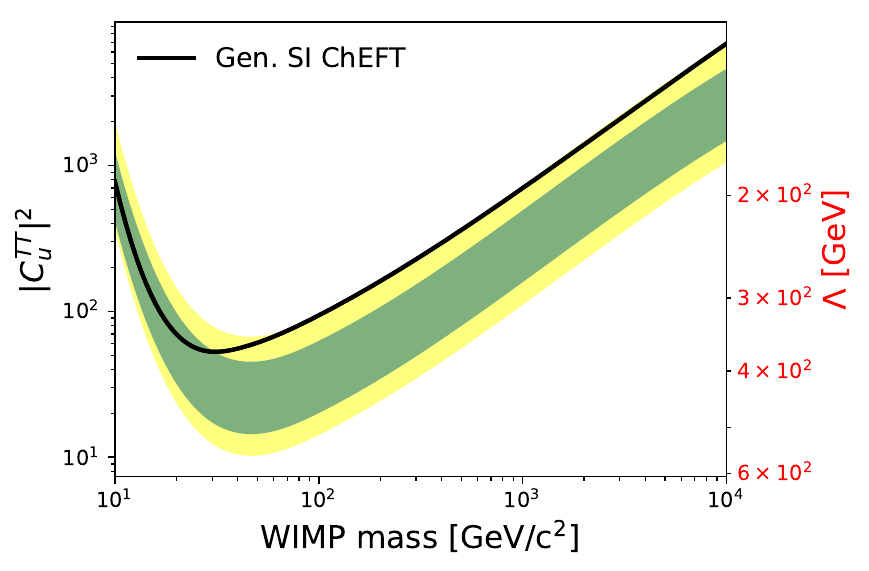}
    \includegraphics[width=3.375in]{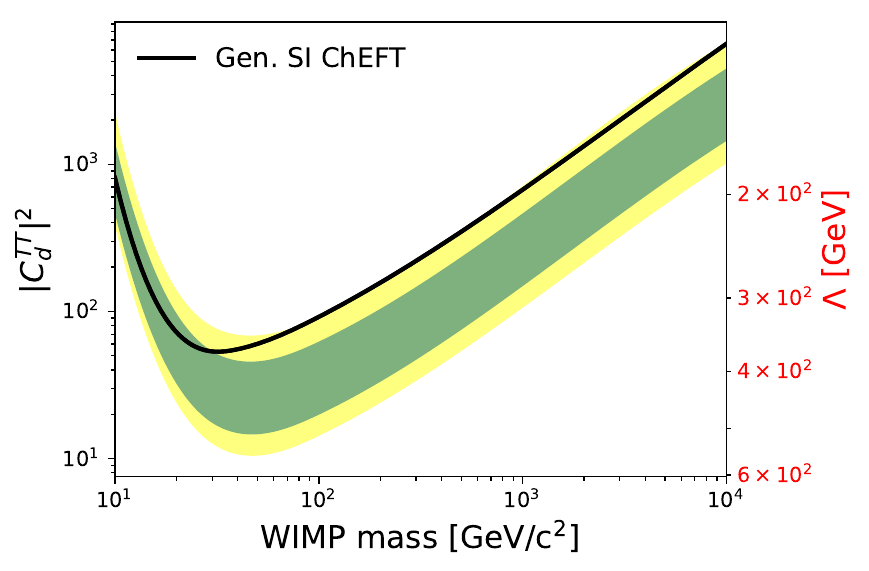}\\
    \includegraphics[width=3.375in]{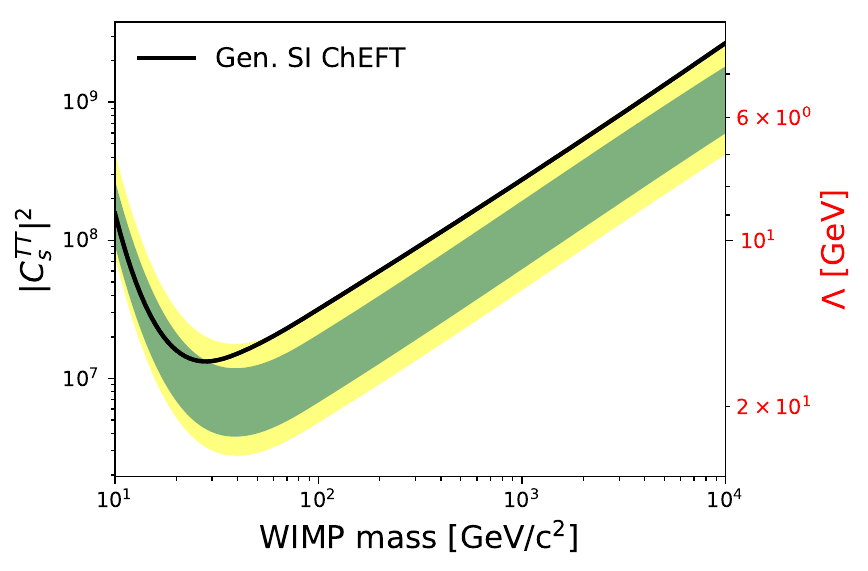}
    \includegraphics[width=3.375in]{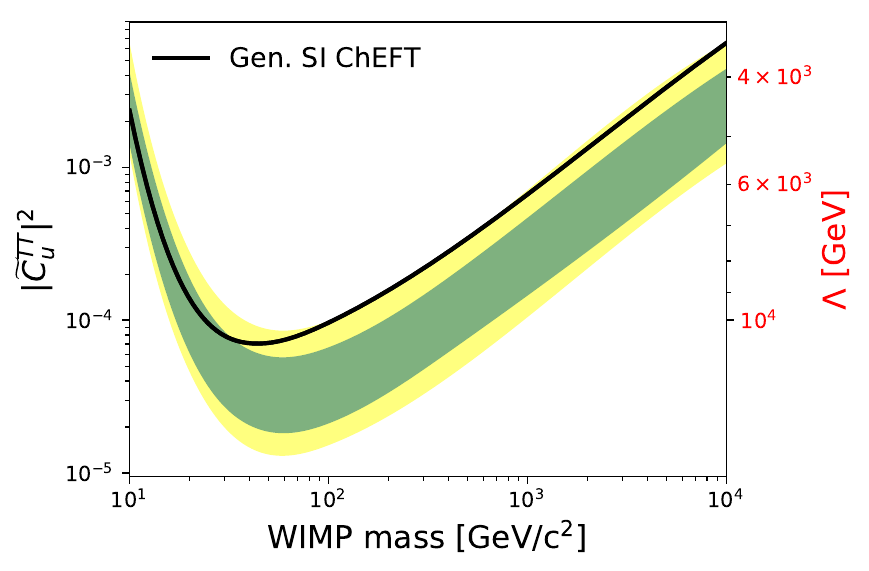}\\
    \includegraphics[width=3.375in]{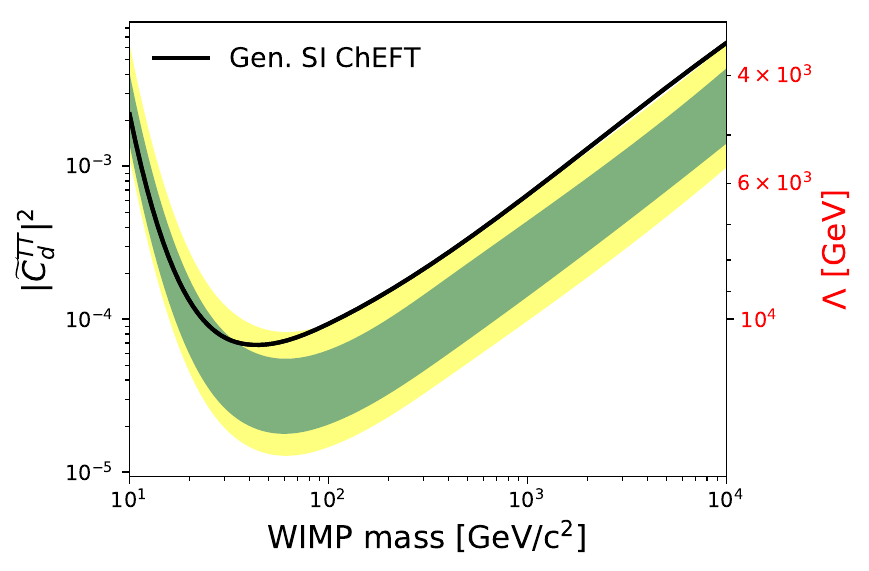}
    \includegraphics[width=3.375in]{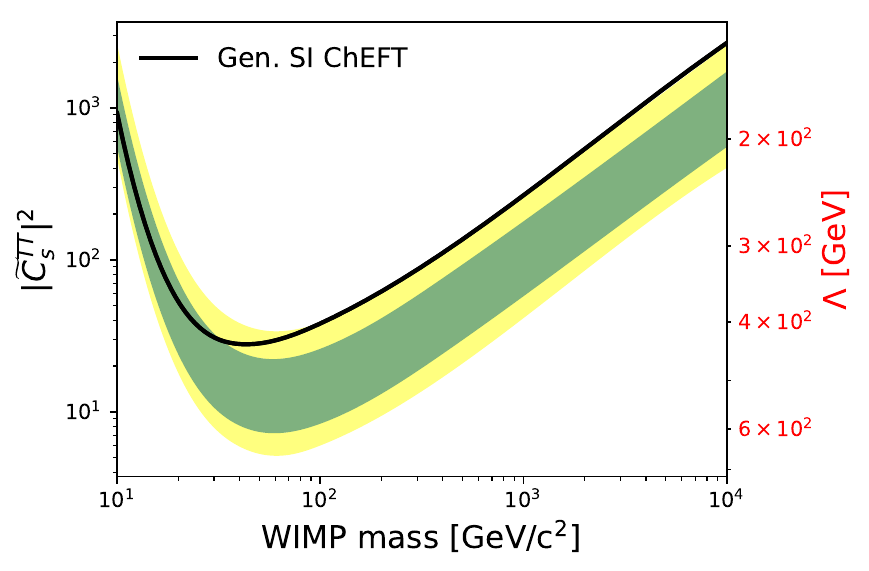}
    \caption{90\,\% CL upper limits (solid black lines) on the Wilson coefficients (left scale) and the physics scale $\Lambda$ (right, inverted scale, in red) for the $TT_u$ (top left), $TT_d$ (top right), $TT_s$ (centre left), $\widetilde{TT}_u$ (centre right), $\widetilde{TT}_d$ (bottom left) and $\widetilde{TT}_s$ (bottom right) couplings in the Generalised SI ChEFT framework, with the $1\,\sigma$ (green) and $2\,\sigma$ (yellow) sensitivity bands.}
    \label{fig:limits_EFT_dim6_2}
\end{figure*}
\begin{figure*}
\centering
    \includegraphics[width=3.375in]{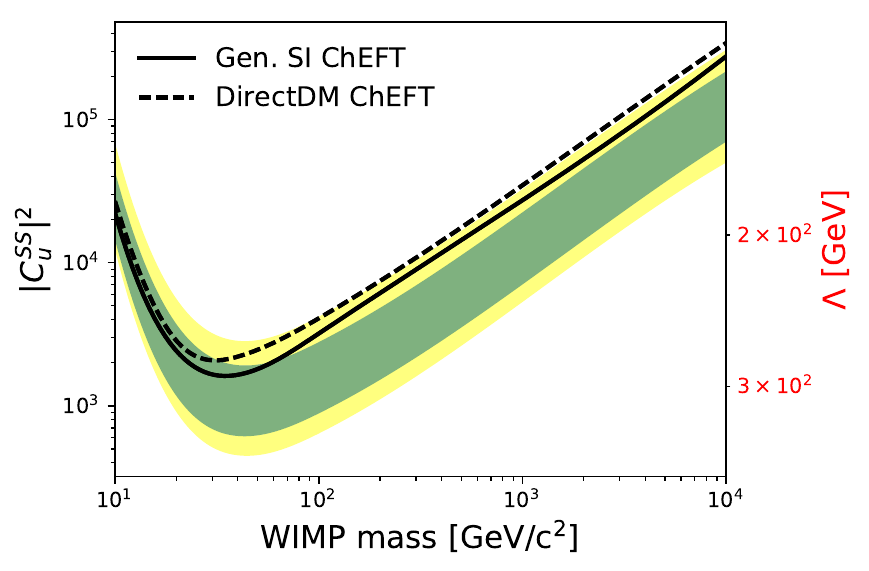}
    \includegraphics[width=3.375in]{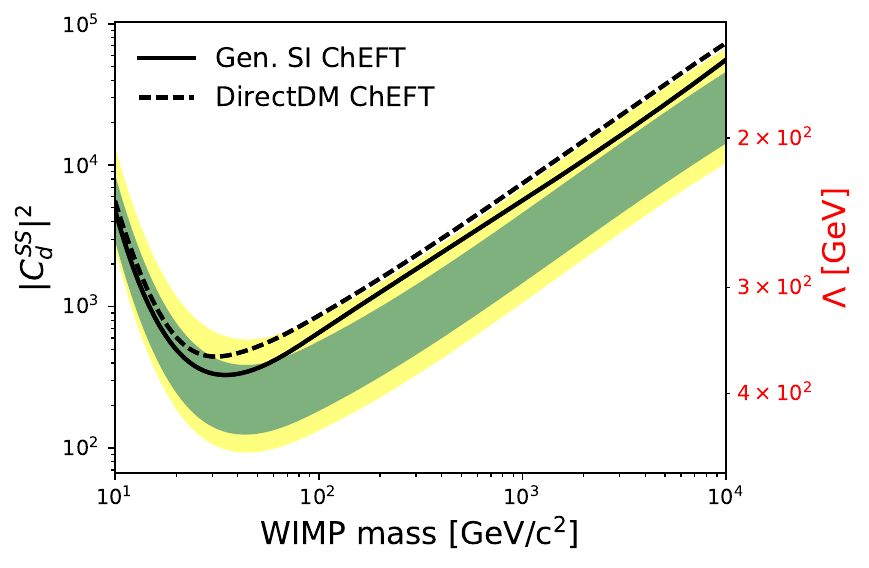}\\
    \includegraphics[width=3.375in]{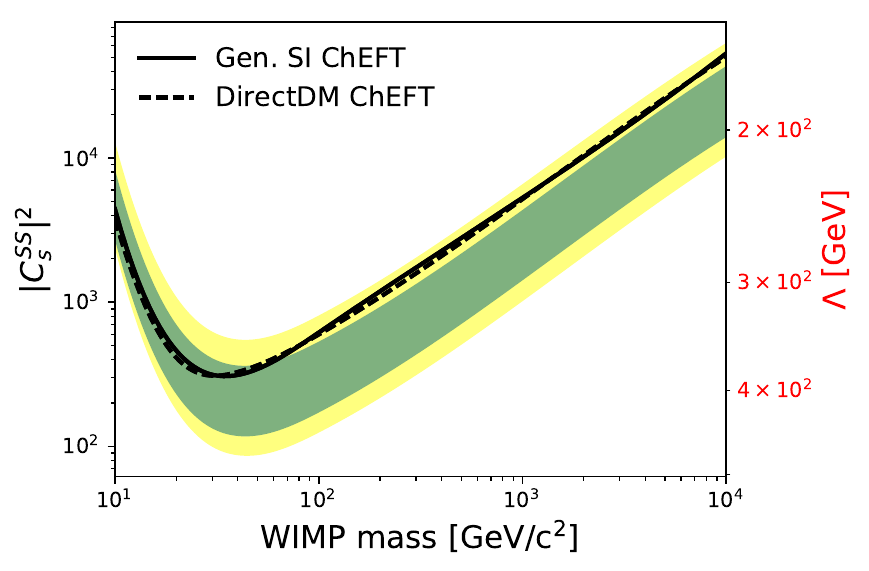}
    \includegraphics[width=3.375in]{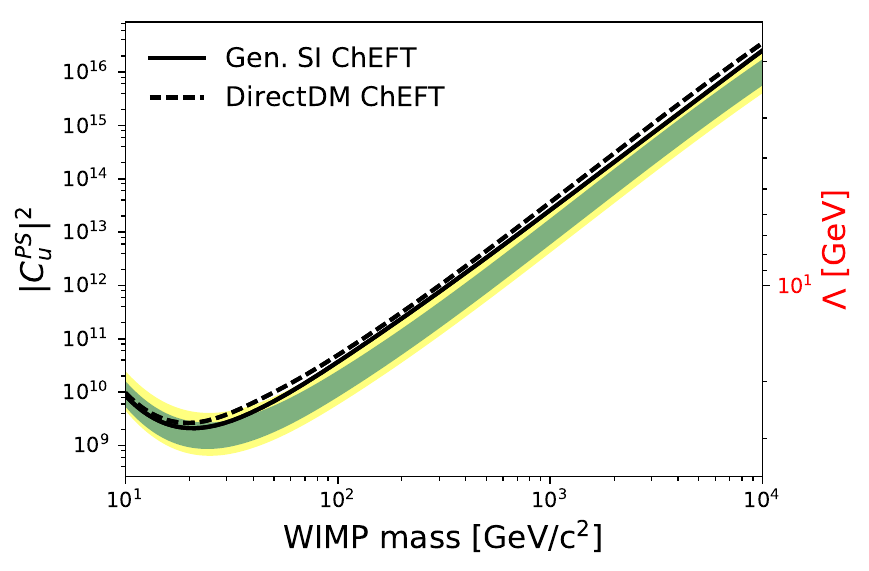}\\
    \includegraphics[width=3.375in]{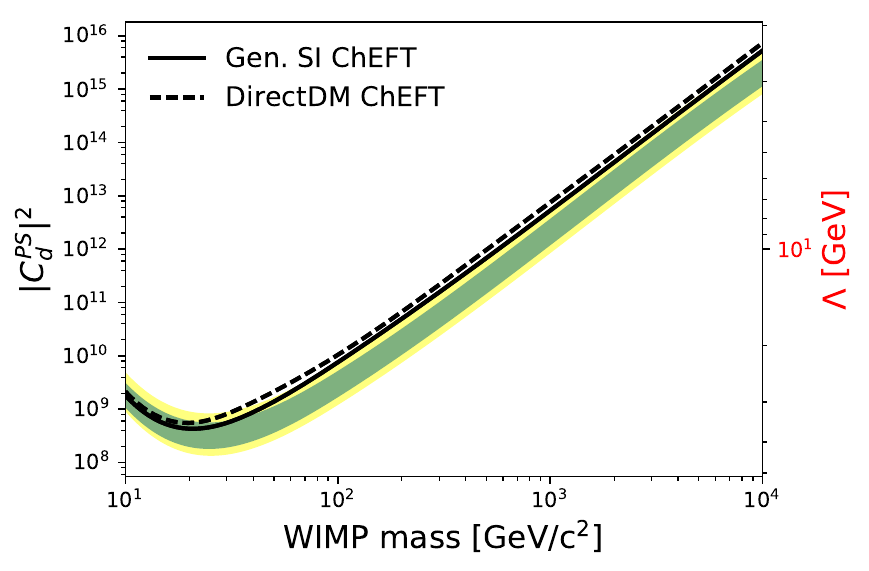}
    \includegraphics[width=3.375in]{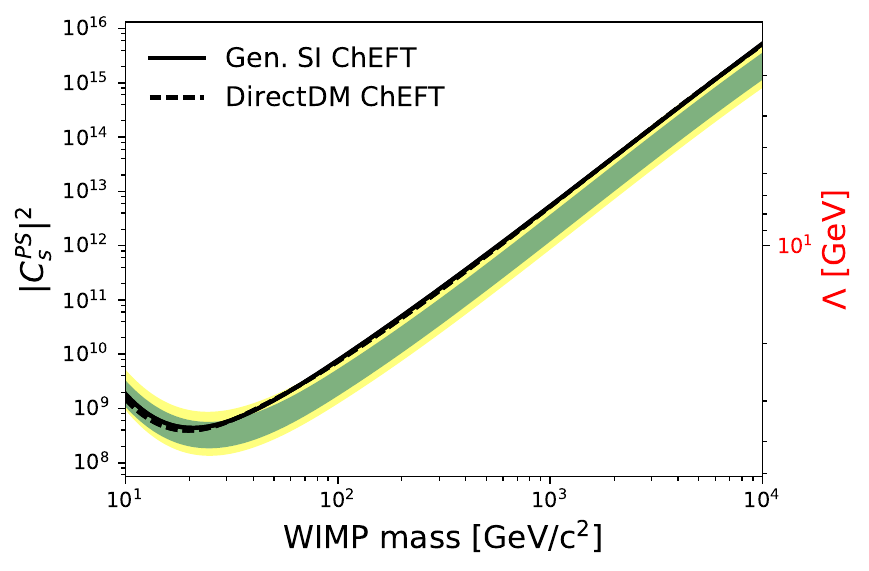}
    \caption{90\,\% CL upper limits (solid black lines) on the Wilson coefficients (left scale) and the physics scale $\Lambda$ (right, inverted scale, in red) for the $SS_u$ (top left), $SS_d$ (top right), $SS_s$ (centre left), $PS_u$ (centre right), $PS_d$ (bottom left) and $PS_s$ (bottom right) couplings in the Generalised SI ChEFT framework, with the $1\,\sigma$ (green) and $2\,\sigma$ (yellow) sensitivity bands. For comparison we show the harmonized 90\,\% confidence level upper limits (black dashed lines) obtained with the DirectDM framework.}
    \label{fig:limits_EFT_dim7}
\end{figure*}
\begin{figure*}
\centering
    \includegraphics[width=3.375in]{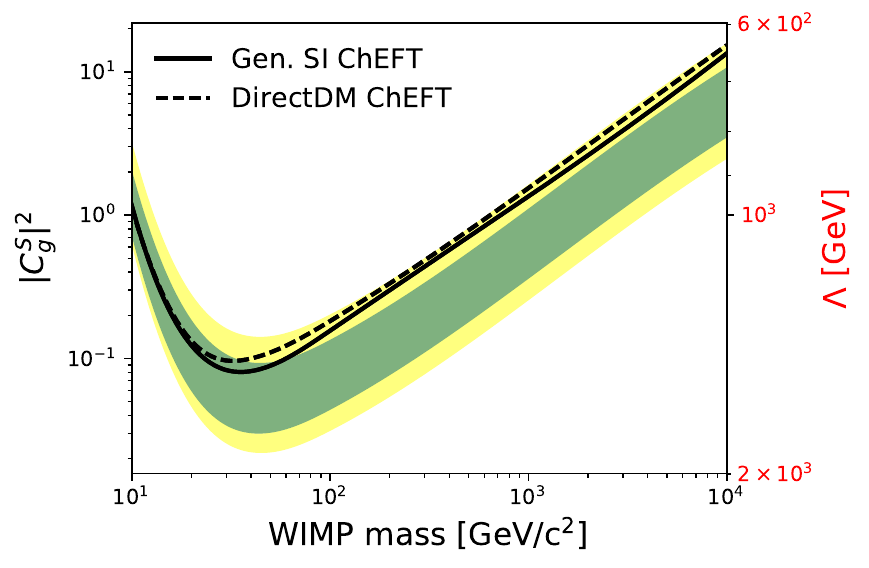}
    \includegraphics[width=3.375in]{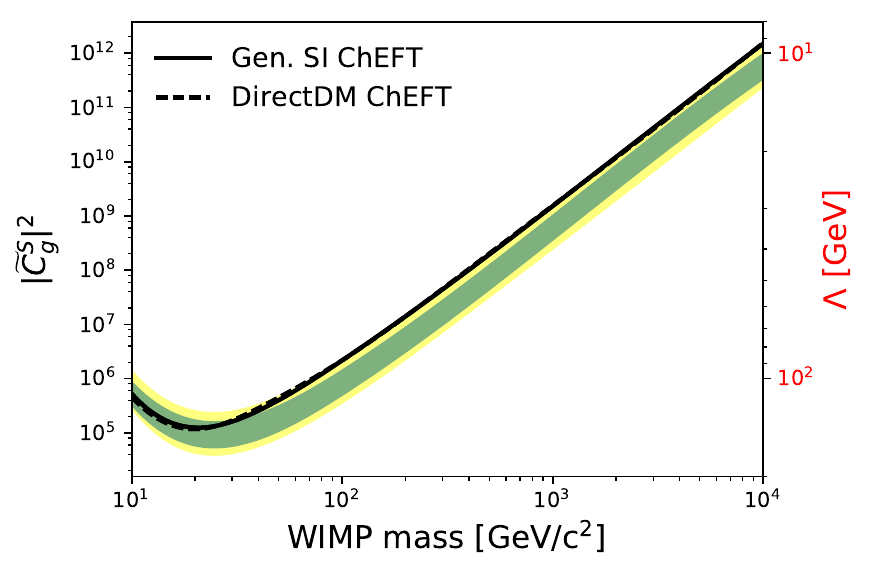}\\
    \includegraphics[width=3.375in]{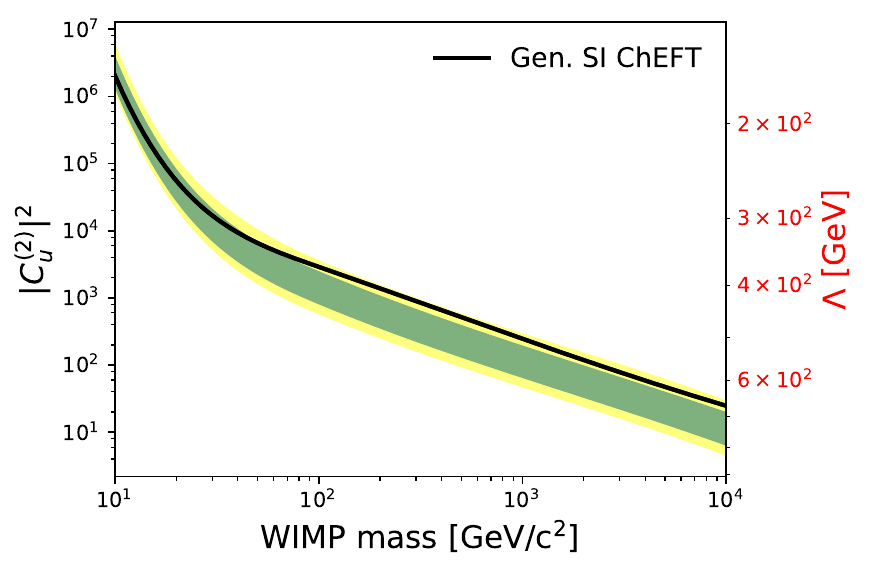}
    \includegraphics[width=3.375in]{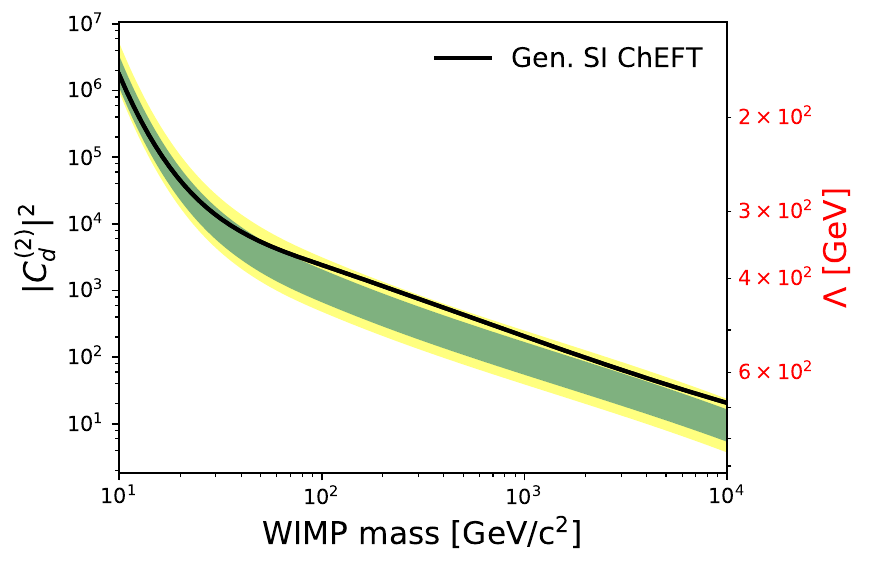}\\
    \includegraphics[width=3.375in]{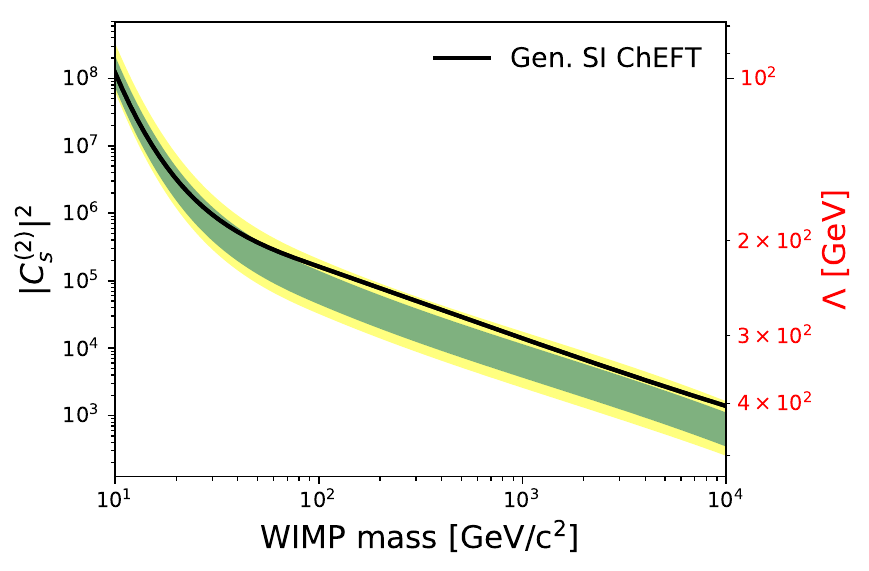}
    \includegraphics[width=3.375in]{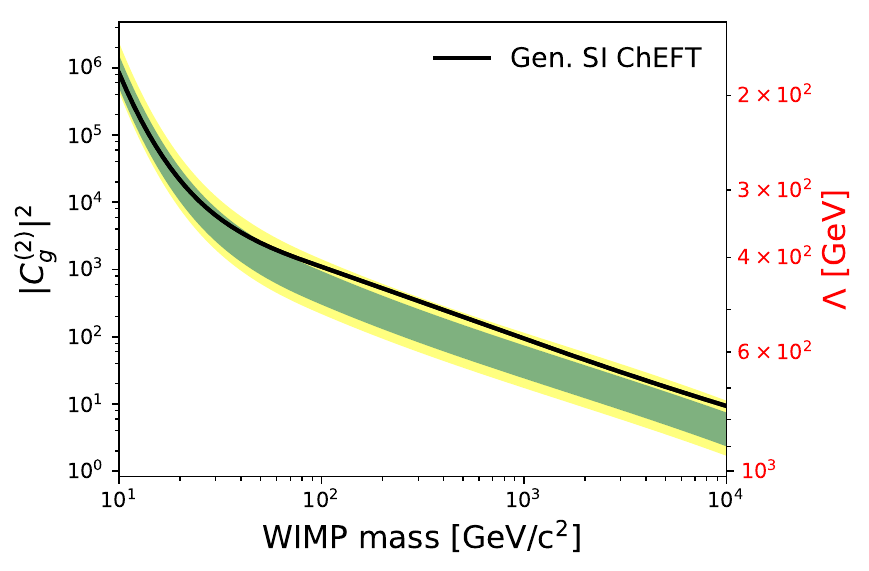}
    \caption{90\,\% CL upper limits (solid black lines) on the Wilson coefficients (left scale) and the physics scale $\Lambda$ (right, inverted scale, in red) for the dimension-seven $S_g$ (top left) and $\Tilde{S}_g$ (top right), and the dimension-eight Spin-2$_u$ (centre left), Spin-2$_d$ (centre right), Spin-2$_s$ (bottom left) and Spin-2$_g$ (bottom right) couplings in the Generalised SI ChEFT framework, with the $1\,\sigma$ (green) and $2\,\sigma$ (yellow) sensitivity bands. For comparison we show the harmonized 90\,\% confidence level upper limits (black dashed lines) obtained with the DirectDM framework for the $S_g$ and $\Tilde{S}_g$ operators.}
    \label{fig:limits_EFT_dim7_2_dim8}
\end{figure*}

A total of 1032 events passed data quality cuts and were included in these analyses. No significant excess was observed in the signal region for any of the models.
The unblinded data set showed three additional events in the NR band region, one in SR0 and two in SR1, statistically compatible with the 1\,\% of the ER band covered by the NR blinding cut in the region. Fig. \ref{fig:signal_models_vs_event_distribution} and Fig. \ref{fig:event_spatial_distribution} show the event distribution and highlight the three new events observed after unblinding.\\
\subsection{Single ChEFT operator results}
For the ChEFT models we obtain the highest discovery significance for the $VV_s$ model for a 70$\,$GeV/c$^2$ WIMP, with a significance of 1.7\,$\sigma$ and a local p-value of 0.043. In Fig.  \ref{fig:limits_EFT_dim5}, \ref{fig:limits_EFT_dim6_1},  \ref{fig:limits_EFT_dim6_2}, \ref{fig:limits_EFT_dim7} and \ref{fig:limits_EFT_dim7_2_dim8} we report the 90\,\% confidence level (CL) limits on the Wilson coefficient for each model at fixed reference value for $\Lambda=1\,$TeV, and the corresponding limit on $\Lambda$ for fixed coefficient value $C_i=1$, against the WIMP mass. These limits are the first experimental limits of this kind from direct detection. For models that produce a response peaked at higher energies, the limits are placed slightly above the $2\,\sigma$ sensitivity band. \\
To harmonize the DirectDM results included in Fig.\,\ref{fig:limits_EFT_dim5}, \ref{fig:limits_EFT_dim6_1}, \ref{fig:limits_EFT_dim7} and \ref{fig:limits_EFT_dim7_2_dim8} with the Generalised SI ChEFT framework, the limits were scaled accordingly by considering the constants in the definition of the operators in \cite{Bishara:2017nnn} and \cite{hoferichter2019nuclear}. In particular the limits of the dipole operators were scaled by a factor $[e^2/(8\pi^2)]^{2}$, and the limits on the $S_g$ and $\tilde{S}_g$ operators were scaled by $(12\pi)^{-2}$. The differences in the limits observed in Fig.\,\ref{fig:limits_EFT_dim7} for $SS_{u,d}$ and in Fig.\,\ref{fig:limits_EFT_dim7_2_dim8} for $S_g$ arise from the contribution of these operators to the two-body interactions considered in the pion matrix elements \cite{hoferichter2019nuclear}, while the difference in the magnetic dipole limit, in Fig.\,\ref{fig:limits_EFT_dim5}, derives from the different contributions to the nuclear response considered, i.e., in the Generalised SI ChEFT framework only the coherent contributions are considered. Other slight differences in the limits computed with the two frameworks can be attributed to the differences of the nuclear responses of the two frameworks. \\

\subsection{Benchmark models results}
\begin{figure*}
    \centering
    \includegraphics[width=2.25in]{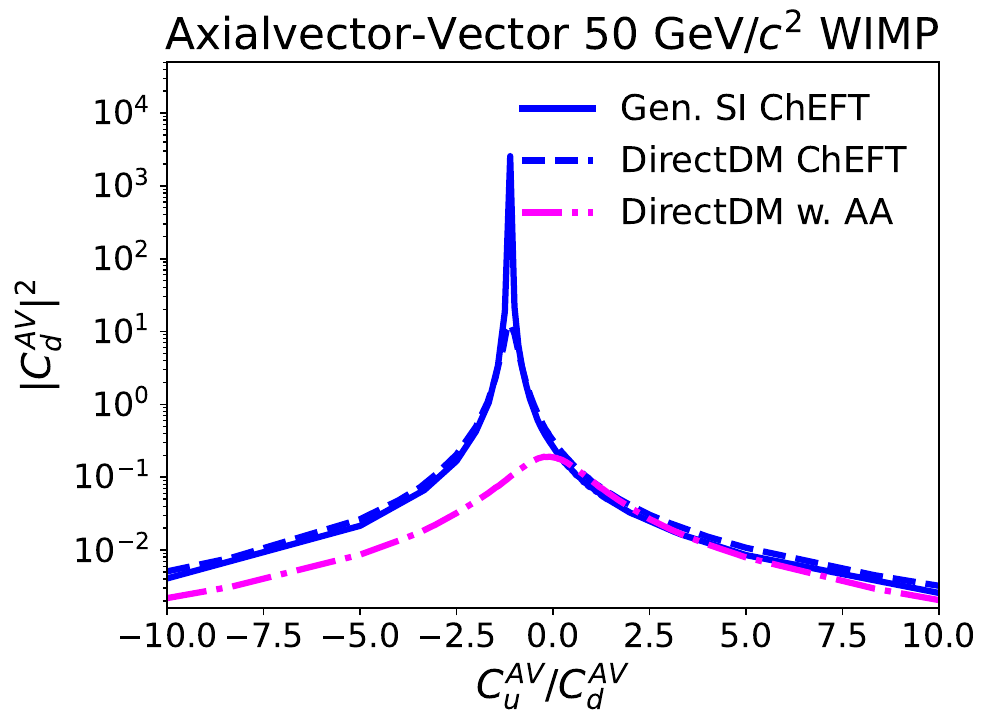}
    \includegraphics[width=2.25in]{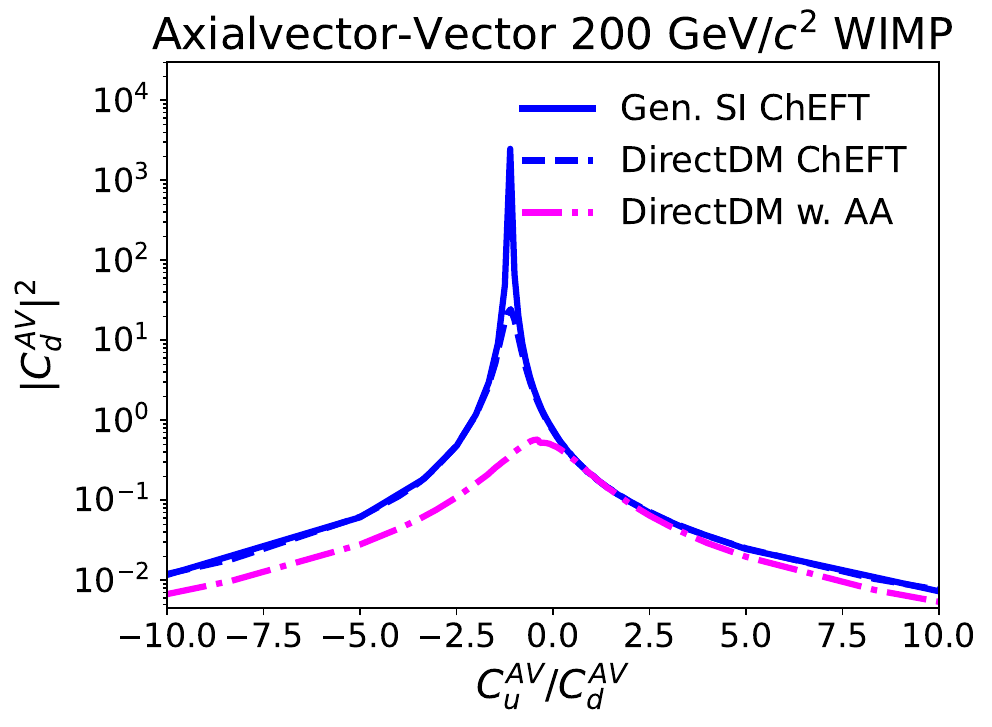}
    \includegraphics[width=2.25in]{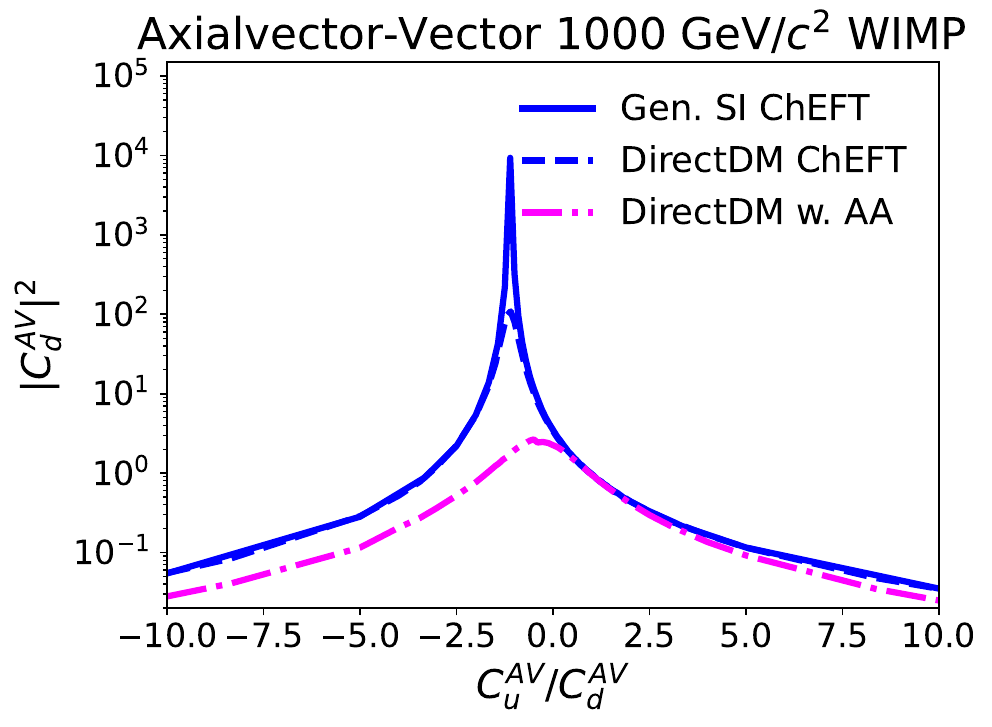}\\
     \includegraphics[width=2.25in]{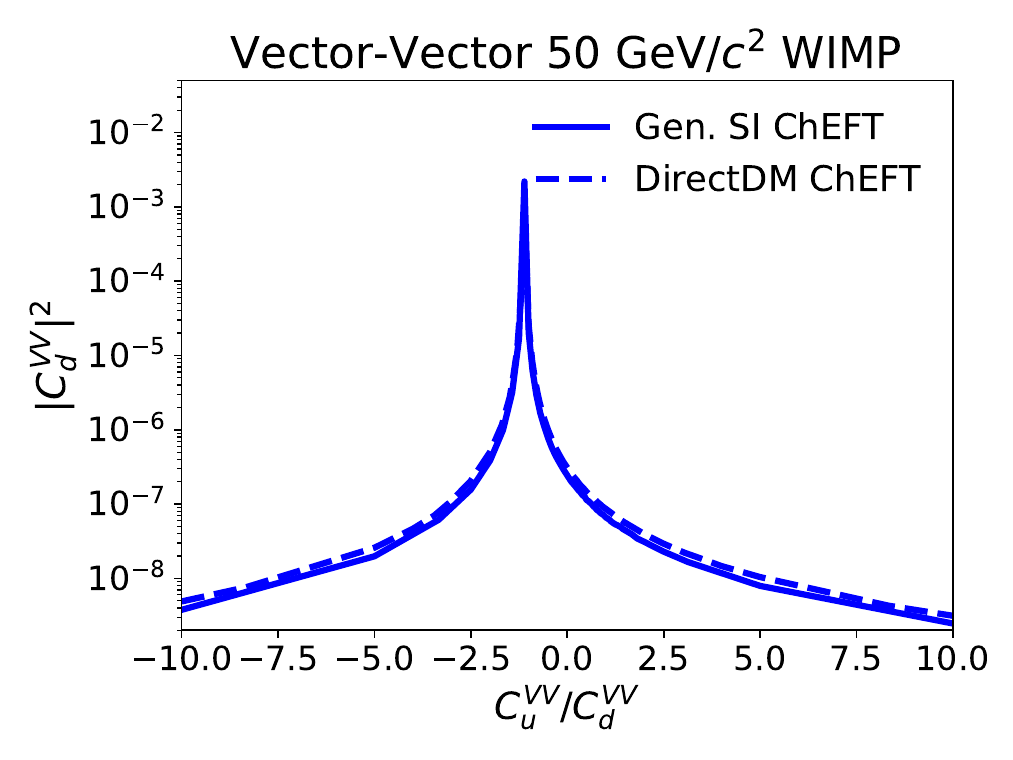}
    \includegraphics[width=2.25in]{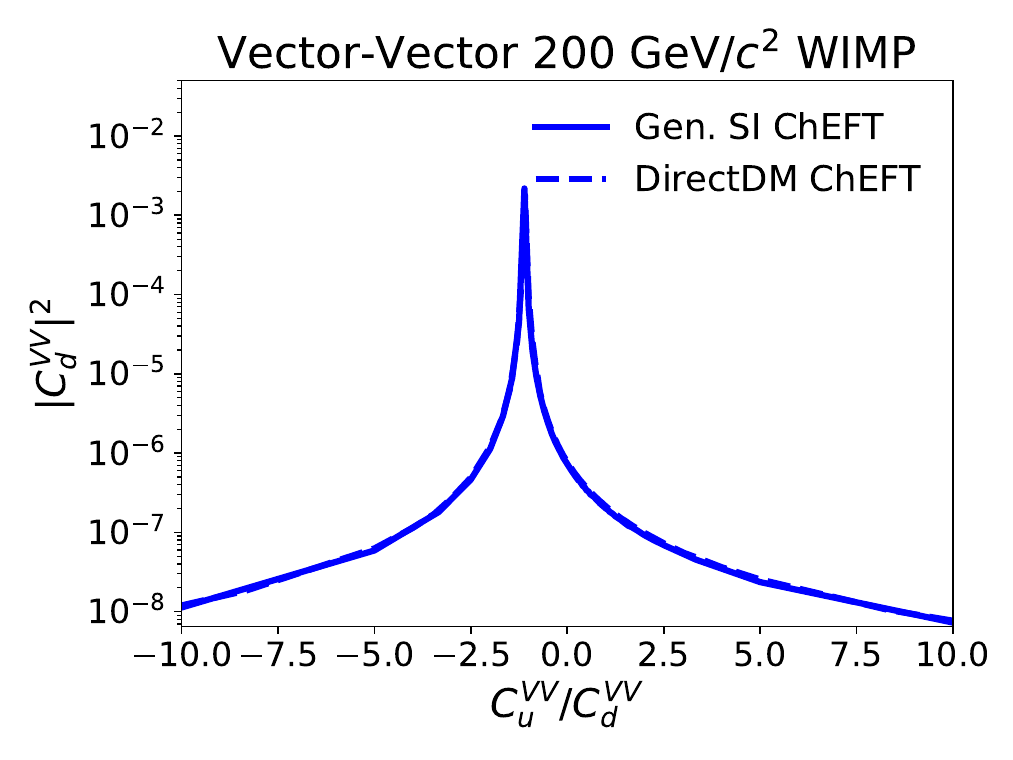}
    \includegraphics[width=2.25in]{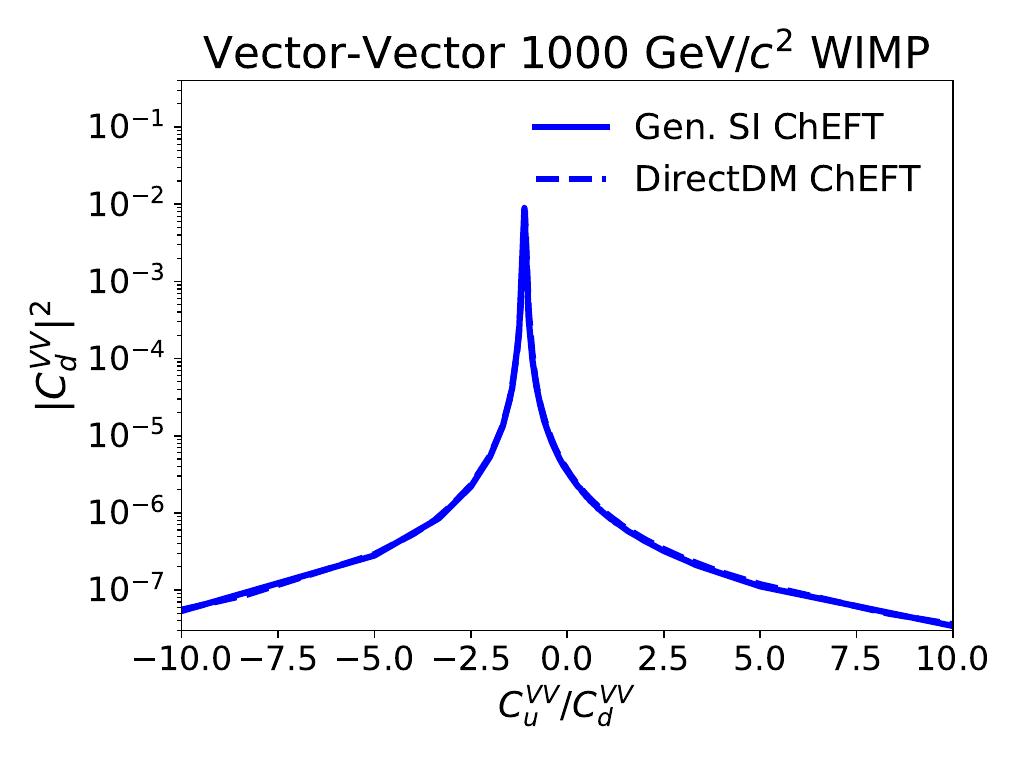}\\
    \includegraphics[width=2.25in]{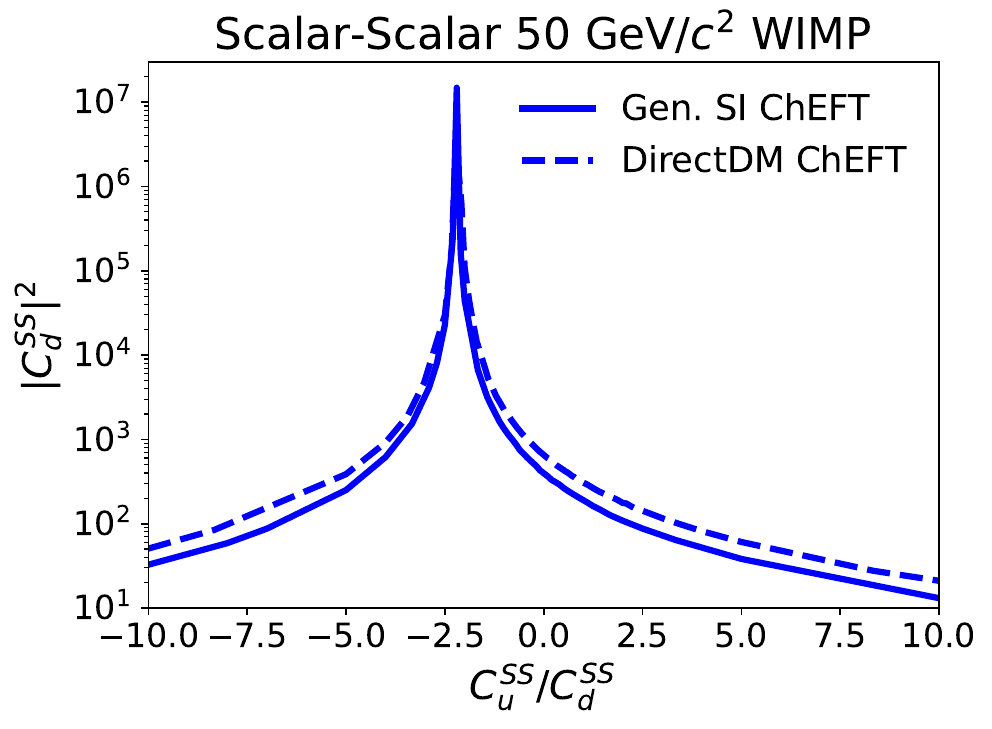}
    \includegraphics[width=2.25in]{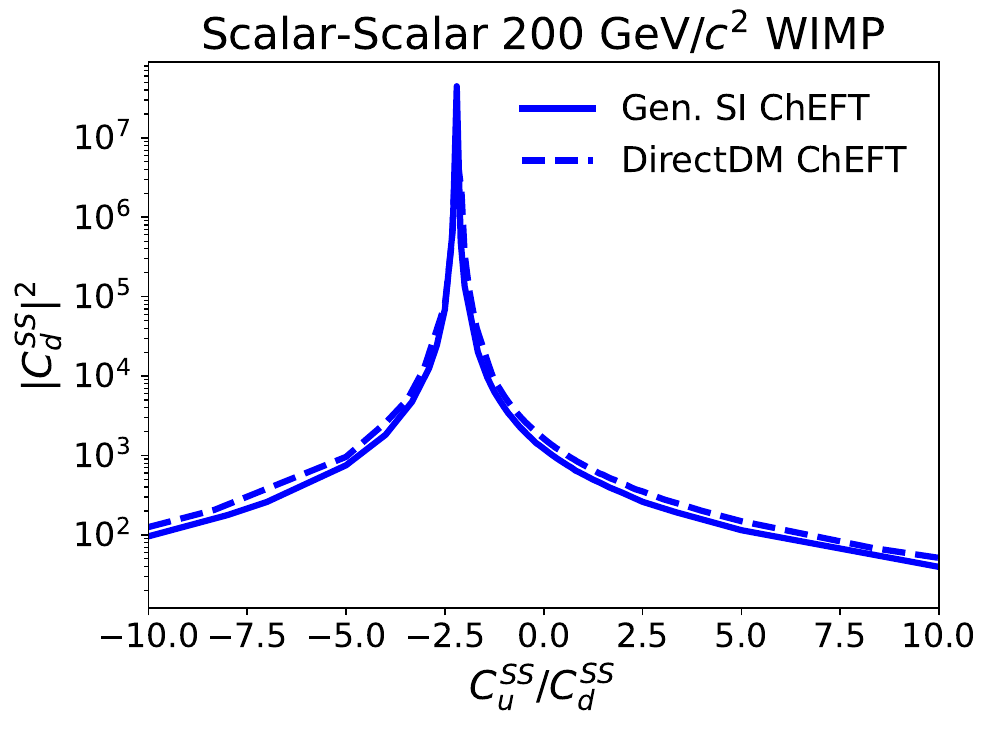}
    \includegraphics[width=2.25in]{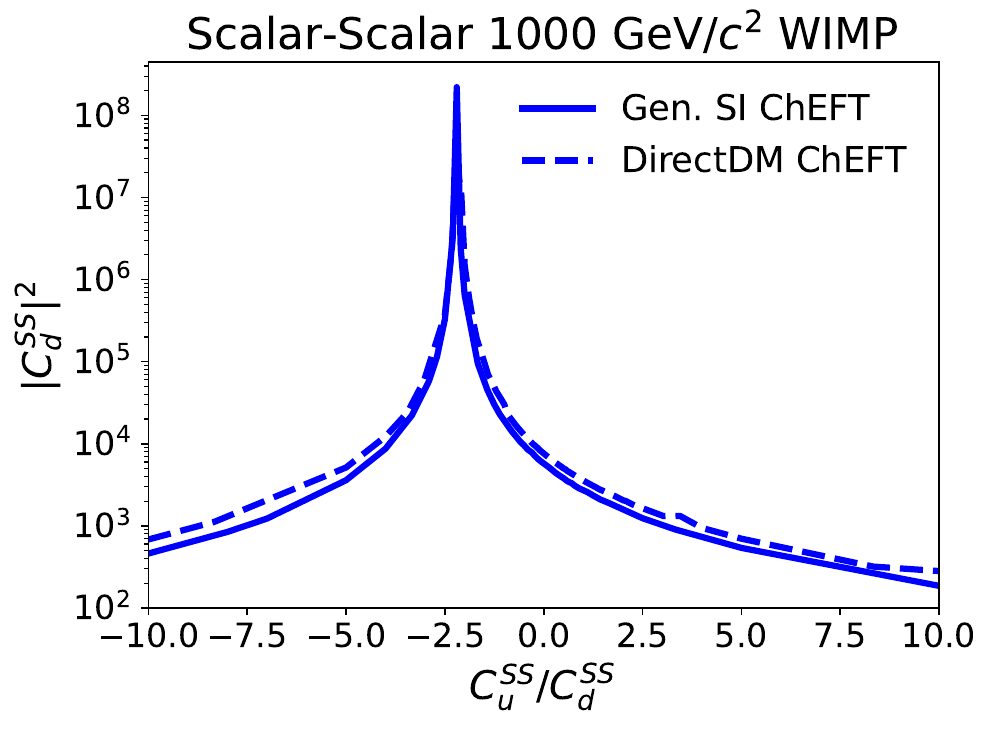}
    \caption{90\,\% CL upper limits on the $down$-quark Wilson coefficients in three benchmark models of WIMP interactions, $AV$ (top row), $VV$ (center row), and $SS$ (bottom row), for three different WIMP masses, 50\,GeV/c$^2$ (left column), 200\,GeV/c$^2$ (center column), and 1000\,GeV/c$^2$ (right column), against the ratio of the $up$ and $down$ reference values of the coefficients. The solid lines represent the limits obtained for models in the Generalised SI ChEFT framework, while the dashed lines are limits obtained from models constructed with the DirectDM framework. For the $AV$ model we show also limits on the $C^{AV}_d$ coefficient when the $AA$ operators contribute with $C^{AA}_d=C^{AV}_d-C^{AV}_u$ (magenta dash-dotted lines), computed for models constructed with the DirectDM framework.}
    \label{fig:benchmark_models}
\end{figure*}

The 90\,\% CL limits for the benchmark models for the $AV$, $VV$ and $SS$ interactions are shown in Fig. \ref{fig:benchmark_models} for three different masses, where we choose to plot the limit on the $C_d^a$ coefficient against the set ratio of $r=C_u^a/C_d^a$. They all show an interference point in the case of isospin-breaking interactions that worsens the limit by 3-4 orders of magnitude for the $AV$ and $VV$ models, while for the $SS$ model it worsens the limit by up to 6 orders of magnitude. For the $AV$ and $VV$ models the interference point is in the region of $r\simeq -1.1$, while for the $SS$ model the interference happens around $r\simeq-2.2$. In the case of the $AV$ model, when the $AA$ contribution is present, the interference effect disappears almost completely.\\
We demonstrate that while it is quite possible to avoid the stringent SI limits by allowing for isospin violation, it requires very fine tuning of the coupling terms. There is only a narrow range of ratios where the interference results in a significantly worse limit. This is largely due to the opposing effect of the interference on protons and neutrons. Other studies have shown similar cancellation effects when investigating isospin-breaking DM interactions \cite{yaguna2017isospin,kelso2018directly}. The methods and results however are not directly comparable, since we are investigating the effects within a ChEFT framework, and not in a classical interpretation of the DM couplings to neutrons and protons.

\subsection{Inelastic Dark Matter results}
\begin{figure*}
    \centering
    \includegraphics[width=.95\textwidth]{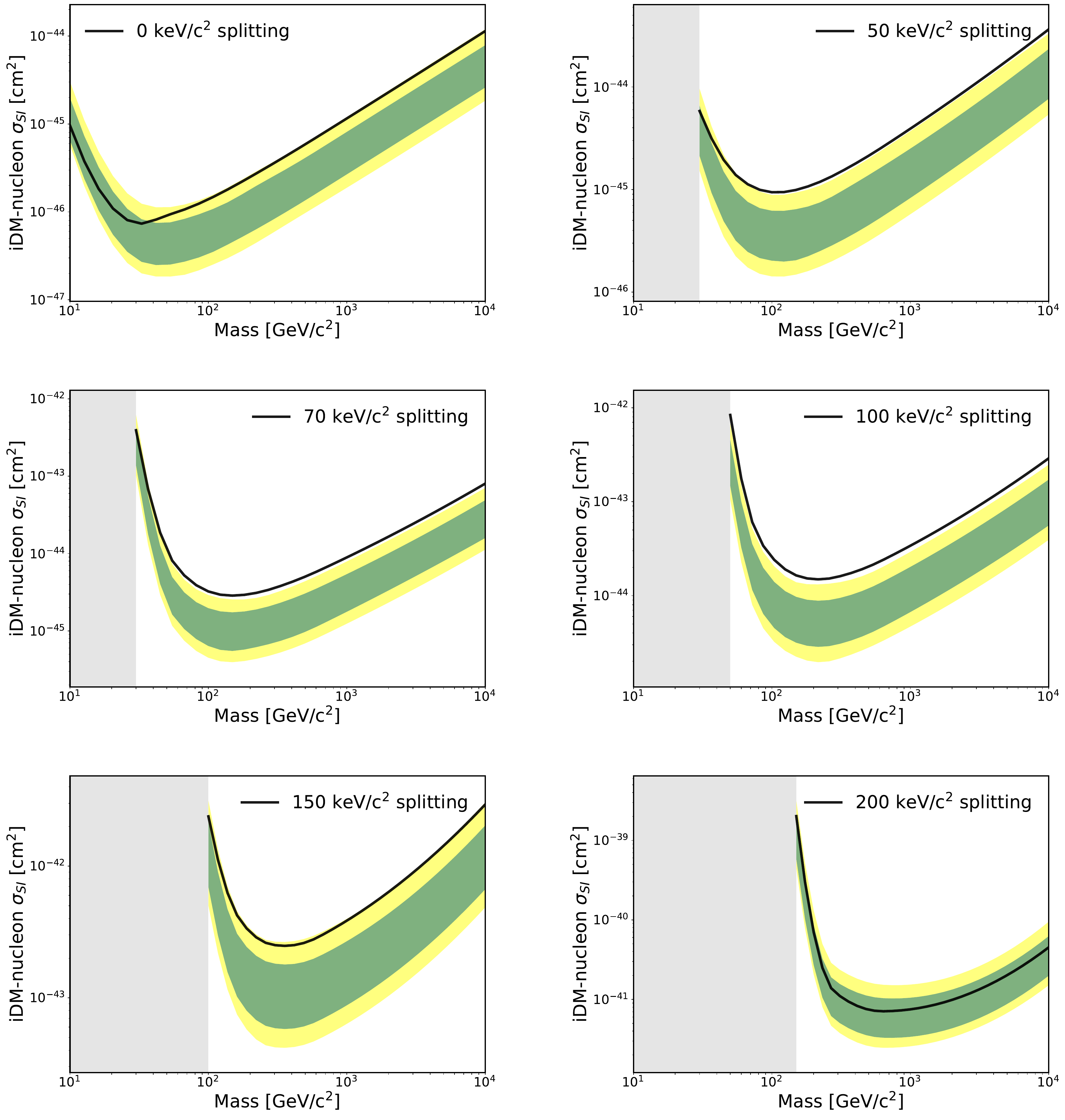}
    \caption{90\% CL upper limits on iDM-nucleon cross sections for selected splittings as a function of mass, spline interpolated. Limit sensitivity bands are shown for 1$\sigma$ (green) and 2$\sigma$ (yellow) of the expected limits under the background-only hypothesis. The grey regions represent masses which were not probed due to the low expected energy transfer.}
    \label{fig:limits_idm_splines}
\end{figure*}

The highest discovery significance for the iDM model was obtained for a mass of 50\,GeV/c$^2$ and splitting of 100\,keV/c$^2$, with a local discovery significance of 1.8\,$\sigma$ and a local p-value of 0.036. The observed p-value, which could be interpreted as slight excess, can be attributed to a slight over-fluctuation in the ER background in the $[26, 27.5]\,$keV$_\text{NR}$ region, resulting in limits being placed on the upper side of the sensitivity band for many models with a peak rate in that energy region. In Fig.\,\ref{fig:limits_idm_splines} we present the 90\,\% CL limits as well as the 1\,$\sigma$ and 2\,$\sigma$ limit sensitivity bands.

\section{Summary}
In this work we perform a comprehensive search of different NR signatures using the combined science data runs SR0+SR1 of XENON1T, for an exposure of 1\,tonne$\times$year, in an extended energy region, up to 100$\,$PE in cS1. After unblinding the extended NR region we observe three new events.
We report the results of WIMP search with the Generalised SI based ChEFT framework providing the first experimental limits on ChEFT couplings up to dimension-eight. We also report limits on the benchmark models of interaction for three different scenarios, the vector mediator for Majorana DM, where the $AV$ operator is the leading contribution; the vector mediator for Dirac DM, where the $VV$ operator contribution is leading; and the scalar mediator for fermion DM case, where the $SS$ term is dominant. In the $VV$ and $SS$ cases we observe a cancellation effect in specific regions of isospin-violating interactions, where the limit worsens by up to four and six orders of magnitude respectively. In the $AV$ case the cancellation disappears for isospin-violating couplings if we correctly consider the rising $AA$ contribution from the above-weak-scale operators. \\
Finally, we report the limits for iDM interactions for mass splittings up to 200$\,$keV/c$^2$, in the case of standard SI interaction. This covers the parameter space the detector is sensitive to. We include these models as they are very localized in the region of the background overfluctuation observed between 20 and 50 PE in the cS1 range and, therefore, represent a good statistical quantification of the maximum local significance of this fluctuation using a physical model. We expect that most of the other possible physical models will produce a similar or lower local significance than the best fit iDM model. 

\section*{Acknowledgements }
We would like to thank M. Hoferichter, A. Schwenk, J. Menéndez, J. Zupan, J. Brod and F. Bishara for providing tools to calculate recoil spectra, the many fruitful discussions and for their guidance and input.\\
We gratefully acknowledge support from the National Science Foundation, Swiss National Science Foundation, German Ministry for Education and Research, Max Planck Gesellschaft, Deutsche Forschungsgemeinschaft, Helmholtz Association, Dutch Research Council (NWO), Weizmann Institute of Science, Israeli Science Foundation, Binational Science Foundation, Fundacao para a Ciencia e a Tecnologia, R\'egion des Pays de la Loire, Knut and Alice Wallenberg Foundation, Kavli Foundation, JSPS Kakenhi and JST FOREST Program in Japan, Tsinghua University Initiative Scientific Research Program and Istituto Nazionale di Fisica Nucleare. This project has received funding/support from the European Union’s Horizon 2020 research and innovation programme under the Marie Sk\l{}odowska-Curie grant agreement No 860881-HIDDeN. Data processing is performed using infrastructures from the Open Science Grid, the European Grid Initiative and the Dutch national e-infrastructure with the support of SURF Cooperative. We are grateful to Laboratori Nazionali del Gran Sasso for hosting and supporting the XENON project.

\appendix

\section{Operators analyzed with both the Generalised SI ChEFT and DirectDM}
\label{app:GenSI-DirDM}

The set of operators that appear at leading order for the most common WIMP models that we chose to investigate with both the Generalised SI ChEFT framework and the DirectDM framework are:

\begin{itemize}
    \item the magnetic and electric dipole operators, which correspond to $\mathcal{Q}^{(5)}_{1}$ and $\mathcal{Q}^{(5)}_{2}$ in DirectDM,
    \item the $VV_q$ operators, corresponding to $\mathcal{Q}^{(6)}_{1,q}$ in DirectDM,
    \item the $AV_q$ operators, corresponding to $\mathcal{Q}^{(6)}_{2,q}$ in DirectDM,
    \item the $SS_q$ operators, corresponding to $\mathcal{Q}^{(7)}_{5,q}$ in DirectDM,
    \item the $S_g$ and $\Tilde{S}_g$ operators, corresponding respectively to $\mathcal{Q}^{(7)}_{1}$ and $\mathcal{Q}^{(7)}_{2}$ in DirectDM,
    \item and the $PS_q$ operators, corresponding to $\mathcal{Q}^{(7)}_{6,q}$ in DirectDM. 
\end{itemize}
The main difference between the two frameworks is that the Generalised SI ChEFT considers the contribution from $SS_q$ and $S_g$ to the nuclear pion couplings, while the two-body currents are cannot be captured as effectively in the DirectDM framework, since it relies on the single nucleon NREFT matching and the DMFormFactor package for the nuclear response. Other differences in normalization and computation of the interaction amplitude are found in $\mathcal{Q}^{(7)}_{1}$ and $\mathcal{Q}^{(7)}_{2}$, which differ from the $S_g$ and $\Tilde{S}_g$ squared amplitudes in the Generalised SI framework by a factor $(12\pi)^{-2}$, and in ${\mathcal{Q}}^{(5)}_{1}$ and ${\mathcal{Q}}^{(5)}_{2}$, which differ from the Generalised SI ChEFT magnetic and electric dipoles squared amplitudes by a factor of $[e^2/(8\pi^2)]^{2}$. Besides the normalization factor, the amplitude of the magnetic dipole in the DirectDM framework includes all the contributions to the NREFT operators, $\mathcal{O}_1$, $\mathcal{O}_4$, $\mathcal{O}_5$ and $\mathcal{O}_6$, for which we refer to the definition in \cite{Fitzpatrick_2013,Fitzpatrick:2012ib,Anand_2014}, while the Generalised SI ChEFT magnetic dipole amplitude considers only the coherently enhanced contributions to operators $\mathcal{O}_1$ and $\mathcal{O}_5$ \cite{hoferichter2019nuclear}.\\ 

\begin{figure*}
    \centering
    \includegraphics[width=.3\textwidth]{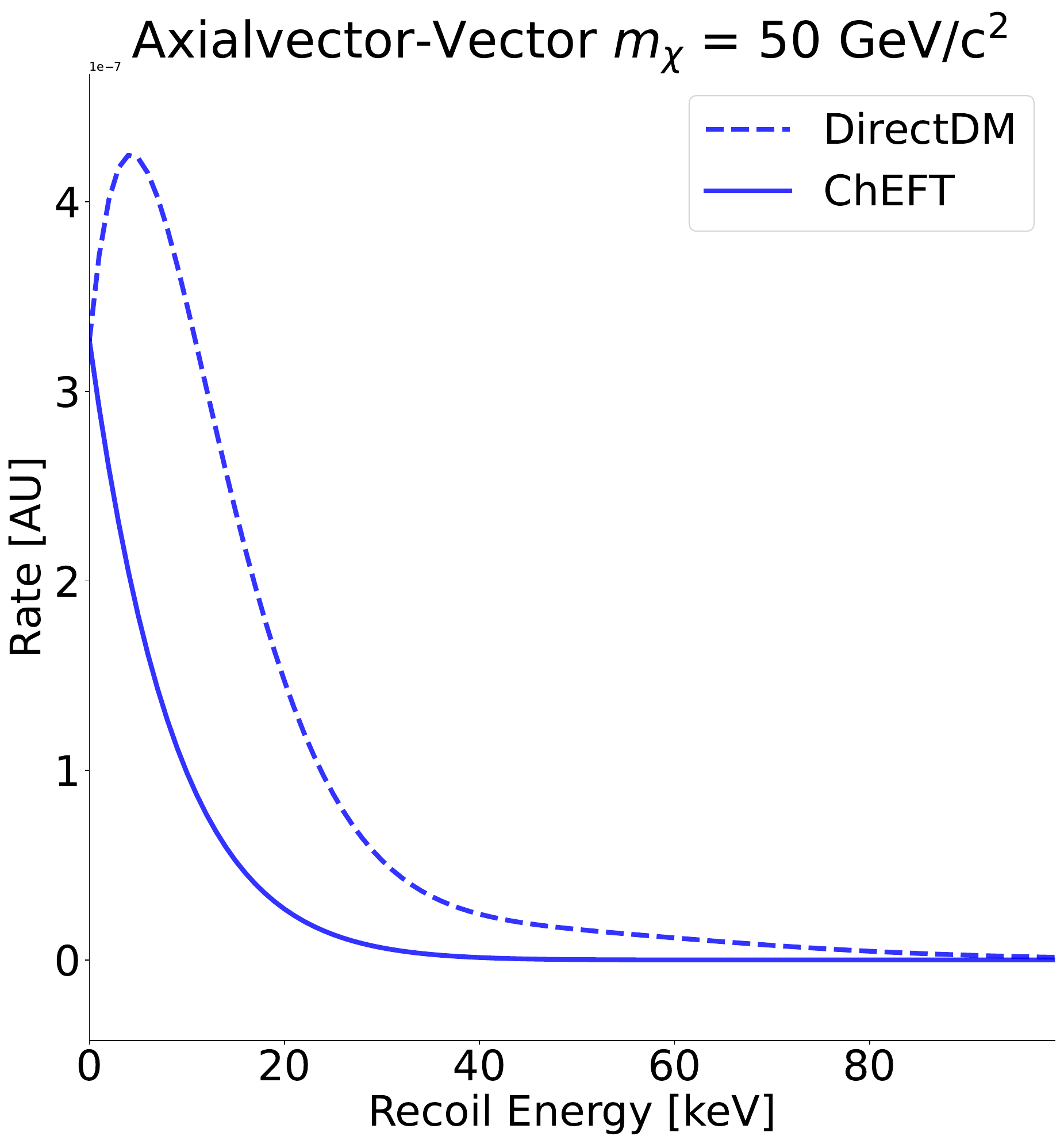}
    \includegraphics[width=.3\textwidth]{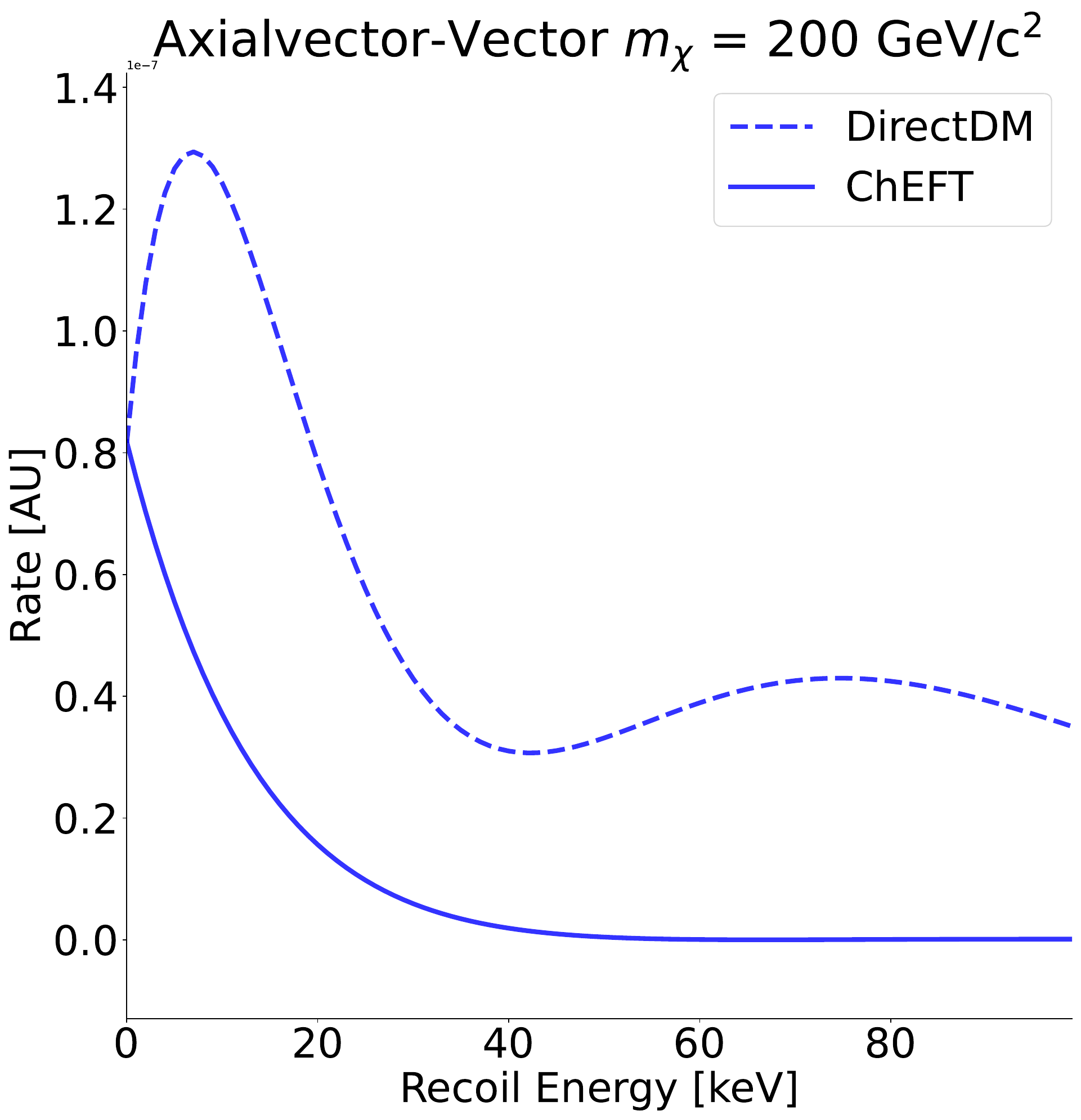}
    \includegraphics[width=.3\textwidth]{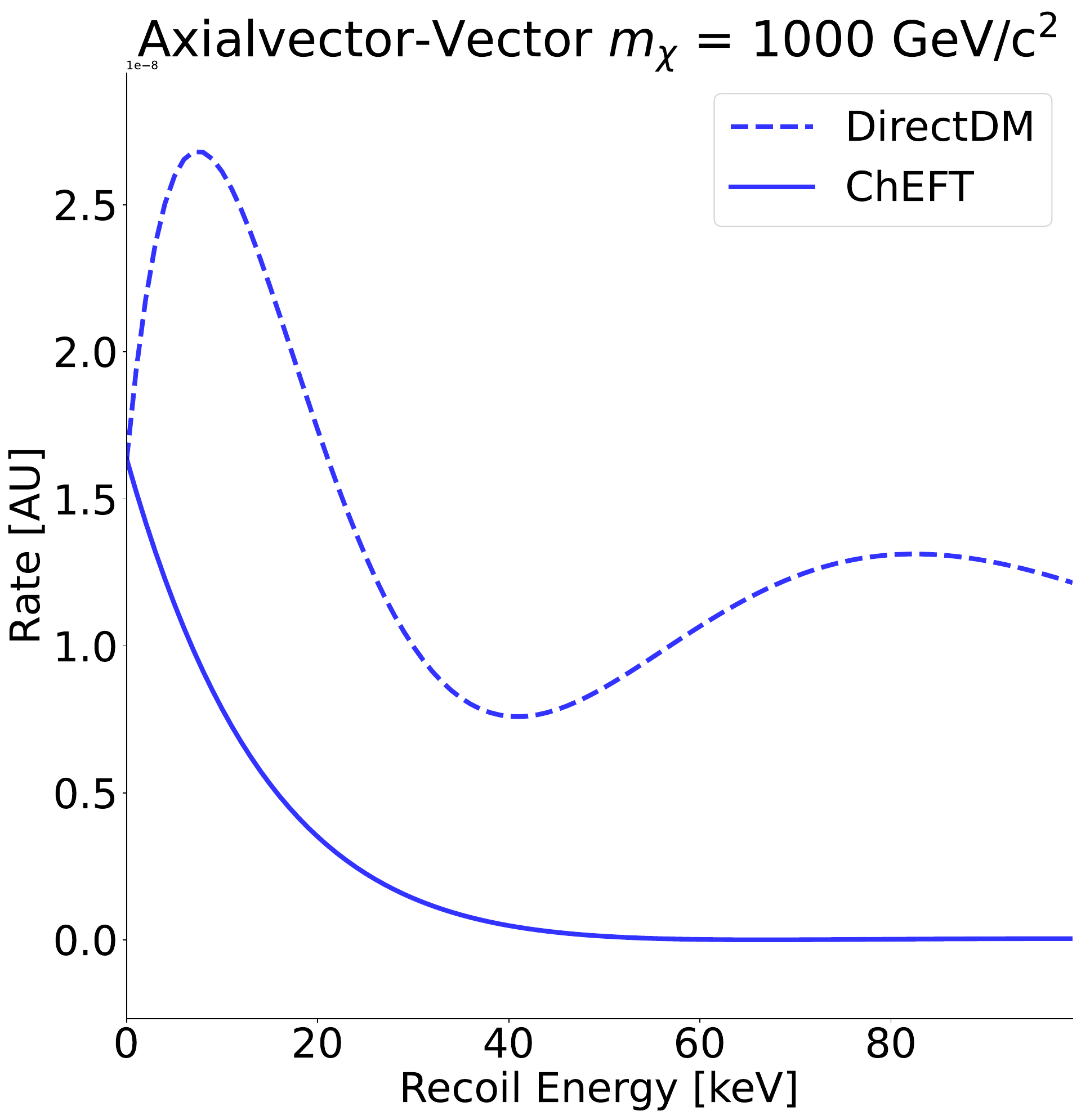}
    \caption{Comparison of the spectra produced by each EFT framework used in this work for the Axialvector-Vector operator with a masses of 50, 200 and 1000 GeV. This operator has the most pronounced differences between the frameworks. }
    \label{fig:compare_dm_frameworks_av_m200gev}
\end{figure*}

\section{Vector mediated Majorana dark matter model}
\label{ap:Majorana}
In the vector-mediated interaction for Majorana DM, the vector current of DM vanishes and the leading contribution comes from the Axial-vector$\otimes$Vector operator \cite{matsumoto2014singlet}, and it is expressed with the Lagrangian of interaction
\begin{equation}
    \mathcal{L}^{AV}_\chi=\frac{1}{\Lambda^2}\sum_{q=u,d} C_q^{AV} \Bar{\chi} \gamma^\mu \gamma_5 \chi\Bar{q}\gamma_\mu q.
    \label{eq:AVmodel}
\end{equation}
In this case, however, due to the $SU(2) \times U(1)$ operators above the weak scale matching onto both $AV$ and $AA$ operators, the $AV_u$ and $AV_d$ contributions cannot be varied independently from the $AA_u$ and $AA_d$. The Wilson coefficients of the $AV$ and $AA$ operators must respect the following relations \cite{bishara2020renormalization}:
\begin{equation}
\begin{split}
    (A \otimes V)_{u} &:  C_{u}^{AV} = \mathcal{C}_{7,1}^{(6)} + \mathcal{C}_{6,1}^{(6)} \\
    (A \otimes V)_d &:  C_{d}^{AV} = \mathcal{C}_{8,1}^{(6)} + \mathcal{C}_{6,1}^{(6)} \\
    (A \otimes A)_u &:  C_{u}^{AA} = \mathcal{C}_{7,1}^{(6)} - \mathcal{C}_{6,1}^{(6)} \\
    (A \otimes A)_d &:  C_{d}^{AA} = \mathcal{C}_{8,1}^{(6)} - \mathcal{C}_{6,1}^{(6)}, 
\end{split}
\end{equation}
where $\mathcal{C}_{i}$ are coefficients of the above-weak-scale operators, i.e. UV operators that are product of DM currents and quark currents.\\
This way, by setting $C^{AA}_u=0$ and $C^{AA}_d = C^{AV}_d - C^{AV}_u$, we retain the freedom to vary $C^{AV}_d$ and $C^{AV}_u$ independently, without further tuning.\\
While an analogous case can be made for the $VV$ and Vector$\otimes$Axial-vector ($VA$) operators, the expected rate of the $VA$ operators is far smaller ($\sim\mathcal{O}(10^{-10})$) than that of the $VV$, and the effect can be safely neglected.

\clearpage
\bibliography{refs}
\end{document}